\documentclass[12pt,preprint]{aastex}

\slugcomment{}

\begin{document}

%% LaTeX will automatically break titles if they run longer than
%% one line. However, you may use \\ to force a line break if
%% you desire.
\shorttitle{3-D Transfer in Hot Star Winds}
\shortauthors{Lobel \& Blomme}

%________________________________________________________________

\title{Modeling Ultraviolet Wind Line Variability in Massive Hot Stars}

%% Use \author, \affil, and the \and command to format
%% author and affiliation information.
%% Note that \email has replaced the old \authoremail command
%% from AASTeX v4.0. You can use \email to mark an email address
%% anywhere in the paper, not just in the front matter.
%% As in the title, use \\ to force line breaks.

\author{A. Lobel$^1$ and R. Blomme$^1$ }

\affil{$^1$Royal Observatory of Belgium, Ringlaan 3, 1180 Brussels, Belgium} \email{alobel@sdf.lonestar.org, Ronny.Blomme@oma.be}

%% Notice that each of these authors has alternate affiliations, which
%% identified by the \altaffilmark after each name.  Specify alternate
%% affiliation information with \altaffiltext, with one command per each
%% affiliation.

%% Mark off your abstract in the ``abstract'' environment. In the manuscript
%% style, abstract will output a Received/Accepted line after the
%% title and affiliation information. No date will appear since the author
%% does not have this information. The dates will be filled in by the
%% editorial office after submission.

\begin{abstract}
We model the detailed time-evolution of Discrete Absorption
Components (DACs) observed in P Cygni profiles of the Si~{\sc{iv}}
$\lambda$1400 resonance doublet lines of 
the fast-rotating supergiant HD~64760 (B0.5 Ib).
We adopt the common assumption that the DACs are caused by Co-rotating Interaction Regions (CIRs)
in the stellar wind.
We perform 3D radiative transfer calculations with hydrodynamic models of the 
stellar wind that incorporate 
these 
large-scale density- and velocity-structures.
We develop the 3D transfer code {\sc Wind3D} to investigate
the physical properties of CIRs with detailed fits to the 
DAC shape and morphology. 

The CIRs are caused by irregularities on the stellar surface that
change the radiative force in the stellar wind. In our hydrodynamic model
we approximate these irregularities by circular symmetric
spots on the stellar surface.
We use the {\sc Zeus3D} code to model the stellar wind and the CIRs, 
limited to the equatorial plane.
We compute a large grid 
of hydrodynamic models and dynamic spectra for the different spot parameters
(brightness, opening angle and velocity).
We demonstrate important effects of these input parameters 
on the structured wind models that determine the detailed DAC evolution.

We constrain the properties of large-scale wind structures with detailed 
fits to DACs observed in HD~64760. 
A model with two spots of unequal brightness and size
on opposite sides of the equator, with opening angles of 
20$\degr$ $\pm$5$\degr$~and 30$\degr$ $\pm$5$\degr$~diameter, 
and that are 20$\pm$5\% and 8$\pm$5\% brighter than the 
stellar surface, respectively, provides the best fit to the observed DACs.
The recurrence time of the DACs compared to the 
estimated rotational period corresponds to spot velocities 
that are 5 times slower than the rotational velocity. 

The mass-loss rate of the structured wind model for HD~64760 does not exceed the rate of the 
spherically symmetric smooth wind model by more than 1\%.     
The fact that DACs are observed in a large number of hot stars
constrains the clumping that can be present in their winds, as substantial
amounts of clumping would tend to destroy the CIRs.
\end{abstract}

%% Keywords should appear after the \end{abstract} command. The uncommented
%% example has been keyed in ApJ style. See the instructions to authors
%% for the journal to which you are submitting your paper to determine
%% what keyword punctuation is appropriate.

%% Authors who wish to have the most important objects in their paper
%% linked in the electronic edition to a data center may do so in the
%% subject header.  Objects should be in the appropriate "individual"
%% headers (e.g. quasars: individual, stars: individual, etc.) with the
%% additional provision that the total number of headers, including each
%% individual object, not exceed six.  The \objectname{} macro, and its
%% alias \object{}, is used to mark each object.  The macro takes the object
%% name as its primary argument.  This name will appear in the paper
%% and serve as the link's anchor in the electronic edition if the name
%% is recognized by the data centers.  The macro also takes an optional
%% argument in parentheses in cases where the data center identification
%% differs from what is to be printed in the paper.


%% \keywords{globular clusters: general ---
%% globular clusters: individual(\objectname{NGC 6397},
%% \object{NGC 6624}, \objectname[M 15]{NGC 7078},
%% \object[Cl 1938-341]{Terzan 8})}

%% From the front matter, we move on to the body of the paper.
%% In the first two sections, notice the use of the natbib \citep
%% and \citet commands to identify citations.  The citations are
%% tied to the reference list via symbolic KEYs. The KEY corresponds
%% to the KEY in the \bibitem in the reference list below. We have
%% chosen the first three characters of the first author's name plus
%% the last two numeral of the year of publication as our KEY for
%% each reference.

\section{Introduction}
Over recent years 
compelling evidence has accumulated
that the mass-loss rates of hot massive stars 
have systematically been overestimated because their winds are not simply steady outflows of 
stellar material, but frequently contain complex density- and velocity-structures. 
There is considerable observational evidence that hot star winds are clumped 
both on small and large length scales. 
The presence of structure significantly influences the mass-loss rate
determinations \citep{Fullerton06, Prinja05, Puls06}.
This in turn is important for models of stellar and galactic evolution.

Hydrodynamic models by 
\citet{Cranmer+Owocki96} showed that large-scale (or coherent) wind structures
in the form of Co-rotating Interaction Regions (CIRs) can 
at least qualitatively explain 
the behavior of Discrete Absorption Components (DACs)
in UV resonance lines of hot stars. 
The DACs are observed to propagate bluewards through the line 
profiles on time scales comparable with the stellar rotation period 
\citep[e.g.][]{Massa+al95a, Prinja98}.
The CIRs are spiral-shaped density and velocity perturbations winding up 
in or above the plane 
of the equator that extend from the stellar surface to possibly several 
tens of stellar radii. 
CIRs can be produced by intensity irregularities 
at the stellar surface, such as dark and bright spots, magnetic loops and 
fields, or non-radial pulsations. The surface intensity 
variations alter the radiative wind acceleration locally, which creates streams of faster and slower material 
through the extended stellar wind. The CIRs are formed where the fast and slow wind regions collide. 

In this paper we investigate to what extent the CIR wind model can explain the 
{\em detailed} wind line variability observed in the B0.5 Ib supergiant HD~64760. 
\citet{Fullerton97} pointed out how exceptional IUE data have made this 
(apparently single) field star ($V$ $\sim$4.24) a key object for studying the origin 
and nature of variability in hot-star winds. For this star there is a substantial 
amount of high-quality IUE observations \citep[MEGA campaign,][]{Massa+al95a}
with unsaturated P Cygni
profiles we utilize to model wind regions where the CIRs dominate the dynamics.
\citet{Fullerton97} propose to clearly distinguish the slowly bluewards shifting 
DACs from a second type of wind line variability called `modulations'. The
latter are most 
pronounced at intermediate blueshifts and drift fast. They are not `discrete' absorption 
components, but are very broad and rather shallow. 
We do not investigate
the modulations here, but we refer to 
\cite{Hamann01}, \cite{Brown+al05} 
and \cite{Krticka+al04} who
provide kinematical models for these modulations.
\citet{Owocki+al94} present a model specifically for HD~64760.

We focus instead on the detailed DAC properties with fully 
hydrodynamic models.
We extend the work of \citet{Cranmer+Owocki96} by using CIR models
to obtain a best fit of the DAC evolution for
HD~64760. 
The star has a high rotational velocity, suggesting we
see it close to equator-on (in what follows, we will assume that
$\sin i=1$): this will considerably 
simplify the modeling of the CIRs and the comparison
to the observations.

\citet{Kaufer+al06} investigate variability observed in optical lines of HD~64760 
and propose a model with perturbations (spots) resulting from the interference of non-radial pulsations
at the base of the wind. This interference pattern does
{\em not} co-rotate with the stellar surface. In this paper 
we show that the shape and evolution of the slow DACs observed in UV wind lines, such as Si~{\sc iv} $\lambda$1395, 
of HD~64760 can correctly be computed with a spot velocity different from 
the stellar surface velocity. The requirement that the CIRs originate from 
spots 
that are not locked onto the surface turns out to be crucial for the 
development of more realistic hydrodynamic models of large-scale wind structures 
in massive hot stars. 

In fitting the HD~64760 observations, we concentrate on the apparent
acceleration of the DAC (i.e.~the evolution of the velocity position
of the flux minimum with time) and the morphology (DAC FWHM evolution
over time). Using a kinematical model,
\cite{Hamann01} showed that the apparent acceleration of DACs is 
always steeper than derived from non-rotating wind models with the same 
velocity law in the radial direction. For spots locked to the surface they 
concluded that the DAC acceleration does not depend at all on 
the stellar rotation rate.   
The present work allows us to check if this conclusion still holds
for spots not locked on the surface.

We implement an advanced 3D radiative transfer code {\sc Wind3D} 
for the detailed modeling of the properties of radiatively driven winds 
(Sect.~\ref{section radiative transfer code} and Appendix). 
It solves the radiation transport problem in three geometric dimensions for winds with
arbitrary density models and velocity fields.
The transfer modeling assumes that the line under consideration is a pure scattering line
(which is an important ingredient of non-LTE) for calculating its detailed shape formed
in a supersonic accelerating wind. 
We investigate the physics of dynamic wind structures by analyzing HD~64760 both 
observationally (Sect.~\ref{obs}) and theoretically (Sect.~\ref{dacfit}). 
We compute an advanced hydrodynamic model and perform radiative transfer calculations in its 
rotating wind. We explore models with structured wind regions that either co-rotate 
with the stellar surface, or rotate slower than the surface (Sect.~\ref{grids}).   
We compare time series of theoretical line profiles (`dynamic spectra') 
with detailed spectroscopic observations of the 
Si~{\sc iv} $\lambda$1395 line variability. In Sect.~\ref{discuss} we discuss some physical properties  
of the large-scale structured wind regions (mass-loss rate and dynamic properties) that 
result from our best fit procedure. We address the important question of how much additional
material the structured wind contains compared to 
spherically symmetric smooth wind models. 
We also compare our results to the \citet{Kaufer+al06} model for
H$\alpha$ variability in HD~64760.
The summary and conclusions are given in Sect.~\ref{concl}.     

\section{Numerical Wind Modeling}

\subsection{3D Radiative Transfer Code Implementation}\label{section radiative transfer code}

We develop the computer code {\sc Wind3D} for the spatial transfer of radiation 
in optically thick resonance lines observed in massive star spectra. Our implementation 
is based on the finite element method described by \citet{Adam90}. 
In the Appendix we present the implementation of the 3D radiative transfer 
scheme by further developing Adam's Cartesian method with three new aspects. 
(i) We considerably accelerate the 3D lambda iteration of the source function with 
appropriate starting values computed with the Sobolev approximation. 
(ii) Since the lambda iteration is the bottleneck of the numerical transfer problem 
we fully parallelize the mean intensity computation. 
(iii) We introduce a new technique that 3D interpolates the converged source function 
to a higher resolution (spatial) grid to solve the final 3D transfer equation for very narrow 
line profile functions. It enables us to resolve small flux variations in the 
absorption portions of very broad unsaturated P Cygni line profiles.                

In the following sections we apply {\sc Wind3D} to model the profiles 
of the Si~{\sc iv} $\lambda$1395 line of HD~64760 and investigate the effects of 
large-scale wind structures on the detailed variability of the emergent line 
fluxes. 
We model only the short-wavelength component of this doublet resonance line.
We can do this as the lines are well separated because the 
terminal velocity ($v_{\infty}$=1500 $\rm km\,s^{-1}$) 
is not too high.
The code could easily be adapted to model both components at the
same time.
In the next Section we test the detailed line formation calculations with 
models of accelerating isothermal winds that incorporate parameterized 3D  
density- and velocity-structures. In Sect.~\ref{section hydro models} we replace the input models with more realistic 
hydrodynamic models that invoke large-scale wind structures in the plane of the equator.               
 
\subsection{3D Parameterized Wind Models}\label{section parameterized models}

Preliminary tests of the {\sc Wind3D} code
were carried out to compare the accuracy of the equidistant 3D rectangular 
grid computations with a simplified radiative transfer (SEI) method.
The latter method \citep[Sobolev with Exact Integration;][]{Lamers+al87} 
is however limited to spherically symmetric winds which is not applicable 
to more realistic asymmetric wind conditions. The SEI method, however, offers a simplified line source 
function, critical for very fast calculations of the detailed shape of resonance lines that form 
in radiatively driven winds. It was therefore temporarily adopted to test the numerical 
accuracy and efficiency of {\sc Wind3D}. Over following code implementation stages the assumptions 
of a smooth and symmetric wind were relaxed to the more realistic conditions of 
structured asymmetric winds. The simplified SEI source function was therefore  
replaced with the fully lambda-iterated line source function.

The {\sc Wind3D} code very efficiently computes the transport of radiation in 
detailed spectral lines formed in extended stellar winds. The FORTRAN code 
is developed for high-performance computers with parallel processing. It has been 
implemented as a fully parallelized (exact) lambda iteration scheme with 
a two-level atom formulation. 
 We also implemented and tested an accelerated lambda iteration scheme (ALI),
but which turned out to not significantly accelerate the overall convergency rates 
for pure scatting lines formed in optically thin
wind conditions (see discussion in Appendix B).
Applications of {\sc Wind3D} to wind conditions that are much more 
optically thick will be given elsewhere.
The model calculations are performed with $71^{3}$ 
grid-points on an equidistant grid. The code lambda-iterates the 3D line source 
function to accuracies below 1\% differences between subsequent lambda iteration 
steps. The local mean intensity integral sums 
80 $\times$ 80 spatial angles over 100 wavelength points 
covering the P Cygni profile.
Next the source function is interpolated to $701^{3}$ grid-points and the 
3D radiative transfer equation is solved to determine the emergent flux.
Both during the lambda iteration and in the final radiative
transfer equation, the intrinsic line profile function is assumed to be
a narrow Gaussian function. 

Tests with spherically symmetric models show that the 1~\%
constraint on the difference between subsequent lambda iteration steps
results in a final source function that is very close to the SEI solution. 
However, more important for our detailed modeling purposes 
is the effect on the resulting line fluxes. We find that these differ by 
less than $10^{-3}$ in the flux normalized absorption portion of the P Cygni 
profile. Lambda iterating to 0.1\% (or 0.01\%) therefore yields 
identical emergent flux profiles, but lengthens iteration times tremendously. 
We emphasize however that the emergent line fluxes we compute strongly depend 
on the very narrow intrinsic line profile width of only
8 $\rm km\,s^{-1}$ we adopt. Our grid of $71^{3}$ points for the 
line source function undersamples the narrow line profile function $\phi_{\nu}$. 
For the actual line flux calculations we therefore have to refine 
the grid to at least $701^{3}$ points, to avoid an undersampling of $\phi_{\nu}$. 
We 3D-interpolate the iterated line source function values to this 
higher-resolution grid (see Appendix~\ref{appendix 3D Radiative Transfer Solution}).
Lower-resolution grids would result in line profiles that are too
(numerically) noisy to compare with the observations.

{\sc Wind3D} has been carefully load balanced for parallel processing with 
the OpenMP programming strategy, and shows excellent scaling properties for multi-threading. 
It accepts arbitrary 3D density models and velocity fields without assumptions of axial symmetry. 
As it uses an observer frame approach, complicated code constructs
to track multiple resonance points are avoided.
{\sc Wind3D} is `fast' and very accurate to trace small variations of local velocity 
gradients and density on line opacities in strongly scattering dominated 
extended stellar winds. 
It currently runs on a 
parallel compute server with 64-bit Itanium-2 1.5 GHz 
microprocessors and a memory architecture allowing for very fast access
to all the memory on the server.
A typical run ($71^{3}$ mesh, 
$80^{2}$ angles, 100 wavelengths) for a single spectral line with `normal' convergence rates of 
the source function takes 
$\sim$300 min of wallclock time using 16 CPUs, followed by another $\sim$300 min to interpolate the 
source function ($701^{3}$ gridpoints), and to compute the dynamic spectrum over 36 lines of sight.
We discuss some memory allocation and parallelization properties in Appendices B \& C.      

We test {\sc Wind3D} with parameterized input models of the wind-velocity and -opacity. 
We consider a $\beta$-velocity law for an isothermal wind with $\beta$=1 
and $R_{*}$=35 $\rm R_{\odot}$.  
The underlying smooth wind reaches a terminal 
velocity of $v_{\infty}$=1600 $\rm km\,s^{-1}$ within the simulation box of 30 $R_{*}$. 
The smooth wind is perturbed with 3D spiraling density structures wound around 
the central star
(the velocities are unchanged from their smooth-wind values).
The source function in the structured model is 3D lambda-iterated 
to equilibrium with the radiation field of the structured wind. The convergency 
of the iterations can be accelerated but our tests show that it is already fast 
and converges within 5 to 8 iterations when the size of the CIRs above and below the 
plane of the equator is limited to 0.5 $R_{*}$ at the outer edge. The converged 
3D line source function is used to solve the transfer problem for 
a uniformly distributed set of sight-angles
in the plane of the equator around the star. In the test input 
models important free parameters for our detailed line profile modeling 
are therefore the properties of the adopted 3D wind structures, such as the number 
of spiral arms, their curvature and width, the height (or flaring angle), 
the density contrast profile throughout the spiral arms, and the inclination 
angle of the observer's line of sight in or above the equatorial plane. 

\begin{figure}  
%\epsscale{1.0}
\plotone{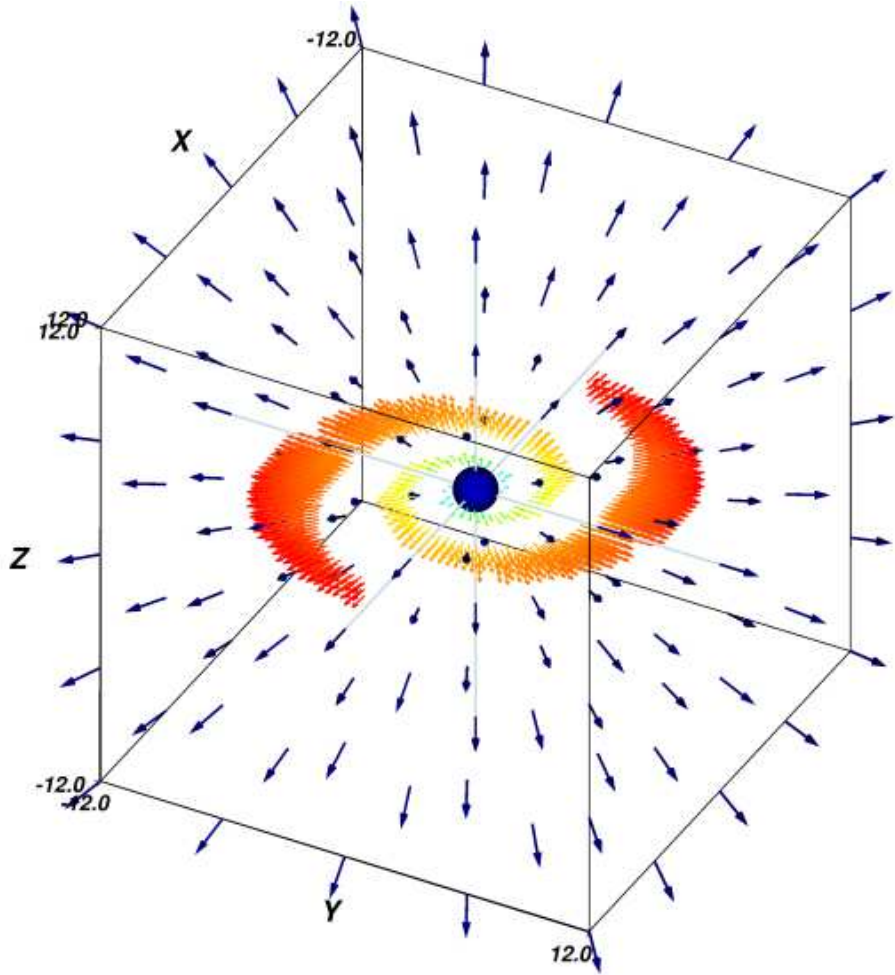} 
\epsscale{1.0} 
\caption{Schematic drawing of a parameterized structured wind model for {\sc Wind3D}. 
The simulation box size is 24 $R_{\star}$, with the hot star at box center. 
The smooth stellar wind is radially symmetric with a $\beta$-power velocity law 
({\it outer arrows}). The local velocities inside the CIRs also assume the $\beta$-law 
of the ambient wind, but the velocity vectors are drawn with much finer spacing
({\it inner set of arrows}). The height of the CIRs above and below the plane 
of the equator is set to 0.5 $R_{\star}$. The wind model rotates over one period and is
viewed by the observer in the plane of the equator for the line profiles in 
Fig.~\ref{fig two DACs}. 
}
\label{fig two CIRs}
\end{figure}

\begin{figure}
%\epsscale{.7}
\vspace*{-5cm}
\plotfiddle{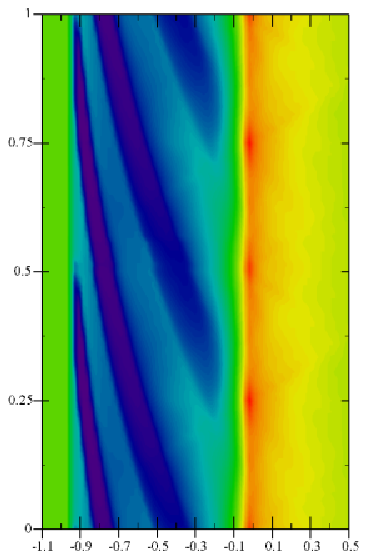}{1.6in}{90.}{260.}{350.}{50}{235}
\vspace*{-2.5cm}
\plotfiddle{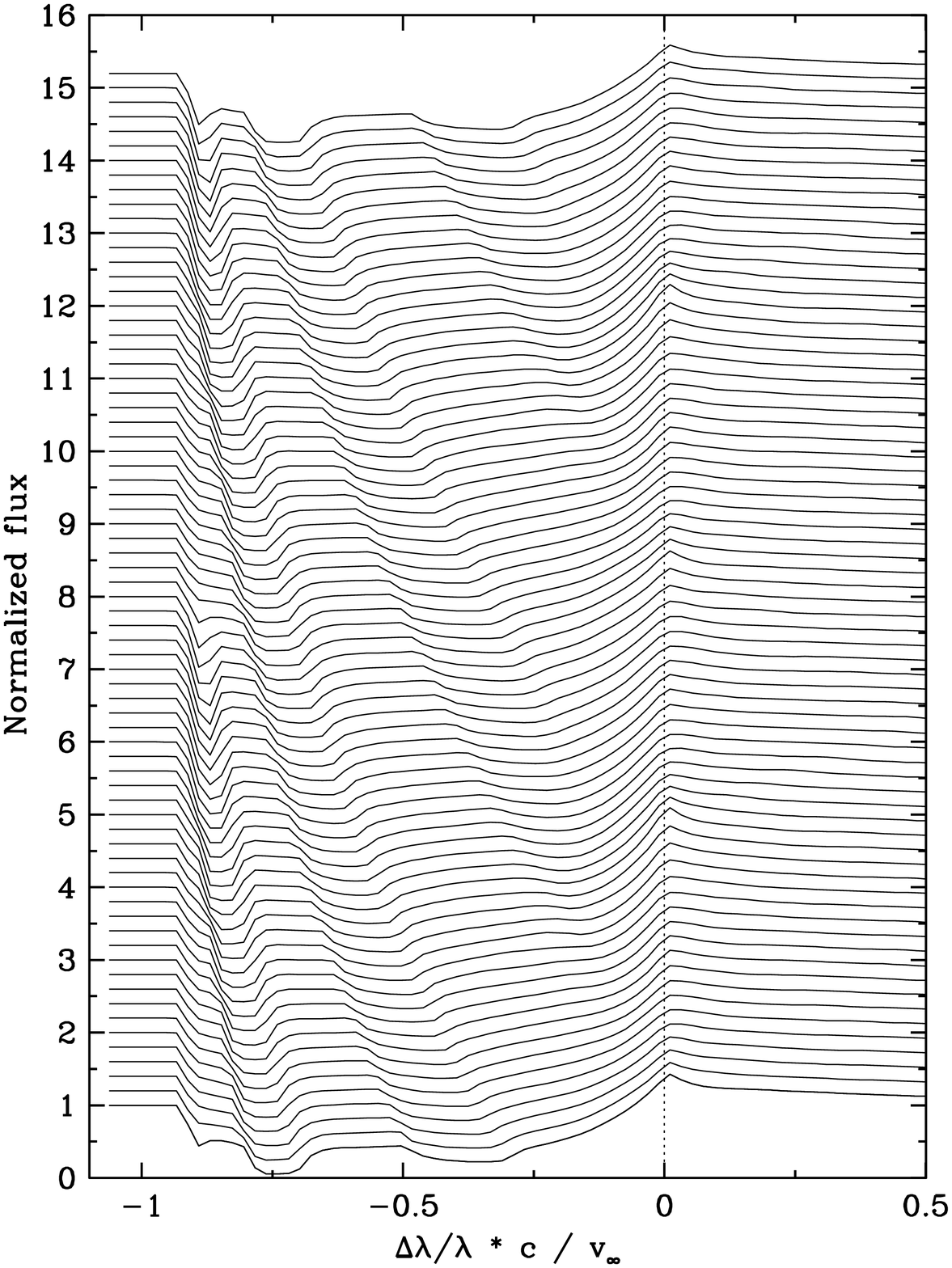}{-3.6in}{90.}{260.}{350.}{50}{225}
\vspace*{-1cm}
\caption{{\bf Left-hand panel:} Spectral time sequence computed with {\sc Wind3D} in the expanding wind of a hot star with two equatorial CIRs shown in 
Fig.~\ref{fig two CIRs}. The theoretical line profiles show two DACs drifting blueward in the unsaturated absorption trough 
of the P Cygni profile (time runs upward over one rotation period). Each spectrum is arbitrarily offset by +0.2 in flux compared to the preceding one.
{\bf Right-hand panel:} The same time sequence of line fluxes shown with grey-scale (color scale in the electronic version) for phases 0 to 1 in the rotational cycle. 
The width of the DACs decreases while shifting blueward. They narrow because the dispersion of wind 
velocities projected in the observer's line of sight in front of the stellar disk decreases 
at larger distances from the surface, while the wind velocity increases to the terminal wind 
velocity. }
\label{fig two DACs}
\end{figure}

Parameterized 3D models of Co-rotating Interaction Regions (CIR) already 
provide comprehensive comparisons to the DAC evolution observed in resonance 
lines of massive hot stars. 
Figure~\ref{fig two CIRs} shows a schematic drawing of a parameterized wind 
velocity grid with two CIRs in the shape of winding density spirals. 
The smooth wind expansion is spherically symmetric with a beta-power velocity law 
({\em outer arrows}). The wind velocities inside the CIRs also assume the beta 
law of the ambient wind and are directed radially ({\em velocity vectors drawn 
with much finer spacing in the equatorial plane}). The widths of both CIRs increase 
outwards with an exponential curvature. The outer edges of the
CIRs are truncated at a maximum radius of $\pm$12 $R_{\star}$.
Figure~\ref{fig two DACs} shows the dynamic spectrum 
computed with {\sc Wind3D} for the structured wind model in Fig.~\ref{fig two CIRs}.
The computed line profiles reveal two (and for certain rotation phases three) 
Discrete Absorption Components (DACs) drifting 
toward shorter wavelengths in the unsaturated absorption trough of the P Cygni line profile 
(time runs upward). The time sequence of these line fluxes is shown in the right-hand panel 
in grey-scale. The line opacity 
{\bf ($\chi^l = \kappa \rho$) }%bf
inside the CIR has been increased by one order of 
magnitude with respect to the surrounding smooth wind opacity. The dynamic spectrum is 
computed for 72 angles of sight in the plane of the equator around the star. 
The Figure shows that the widths of the DACs decrease while they asymptotically drift 
toward shorter wavelengths. The DACs shift toward the blue edge of the P Cygni profile 
because the observer probes regions of increased wind absorption inside the CIR at larger 
distances from the star 
as
the entire CIR structure rotates through the line of sight.  
The widths of the DACs decrease because the range of (radial) velocities 
in the CIR projected in the observer's line of sight (inside the absorbing cylinder in front 
of the stellar disk) decreases at larger distances from the star. At large distances from 
the star a radially expanding spherically symmetric wind has a smaller dispersion of 
radial velocities projected into the line of sight for the wind volume in front of the 
stellar disk. 
In Sect. 4.3.2 we investigate the DAC line formation with advanced hydrodynamic models 
in which both the CIR density and resulting wind velocity gradients turn out to be 
important for the detailed DAC evolution. 

\begin{figure}  
%\epsscale{1.0}
\plotone{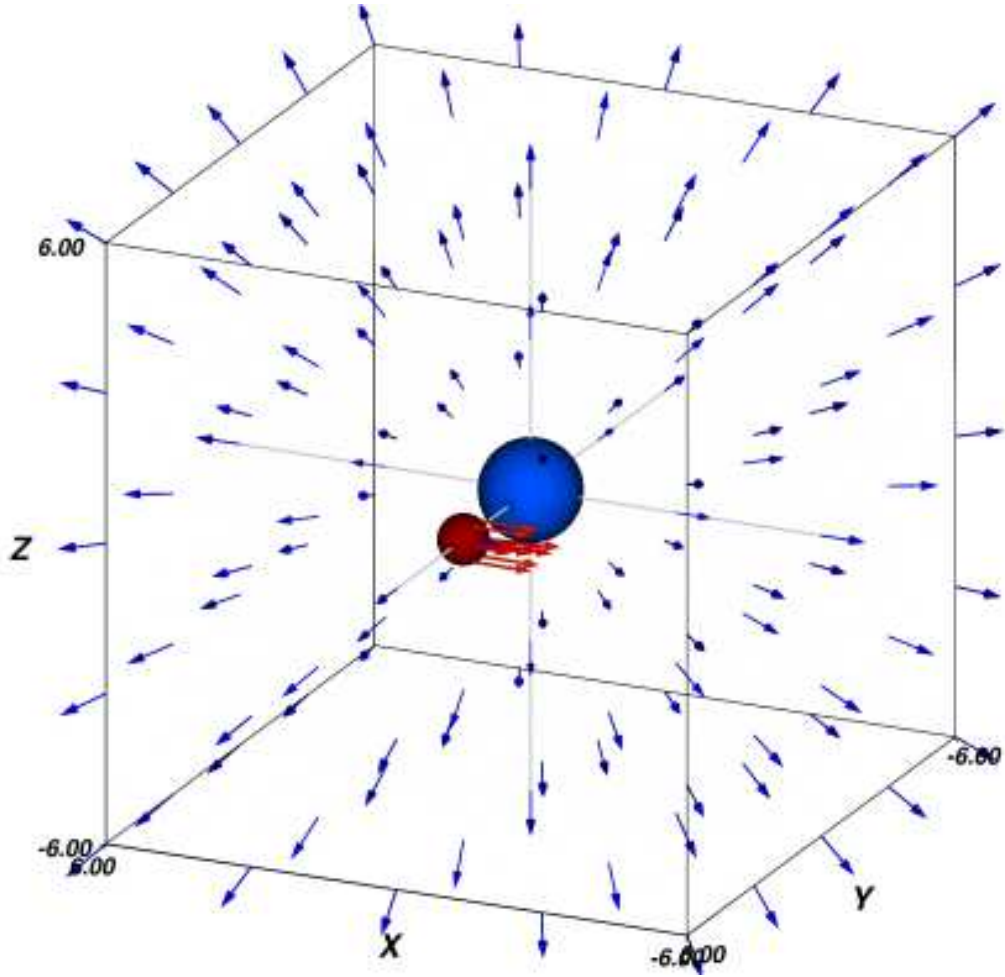} 
\epsscale{1.} 
\caption{Schematic drawing of the wind velocity model with a spherical clump ({\it small front sphere}). 
The blob has a radius of 0.5 $R_{\star}$ and passes at 3 $R_{\star}$ in front of the central hot star 
({\it large sphere}). The size of the simulation box is 12 $R_{\star}$. The blob moves perpendicular 
({\it tangentially drawn arrows}) to the radially accelerating wind ({\it outer arrows}). 
The opacity in the blob is increased by an order of magnitude compared to the ambient wind opacity.
The dynamic spectrum in Fig.~\ref{fig spectrum blob} is computed for 13 lines of sight in the plane of the equator over 
$\pm$45$\degr$ in the front plane of the simulation box.}
\label{fig blob}
\end{figure}

\begin{figure}  
\epsscale{0.8}
\plotone{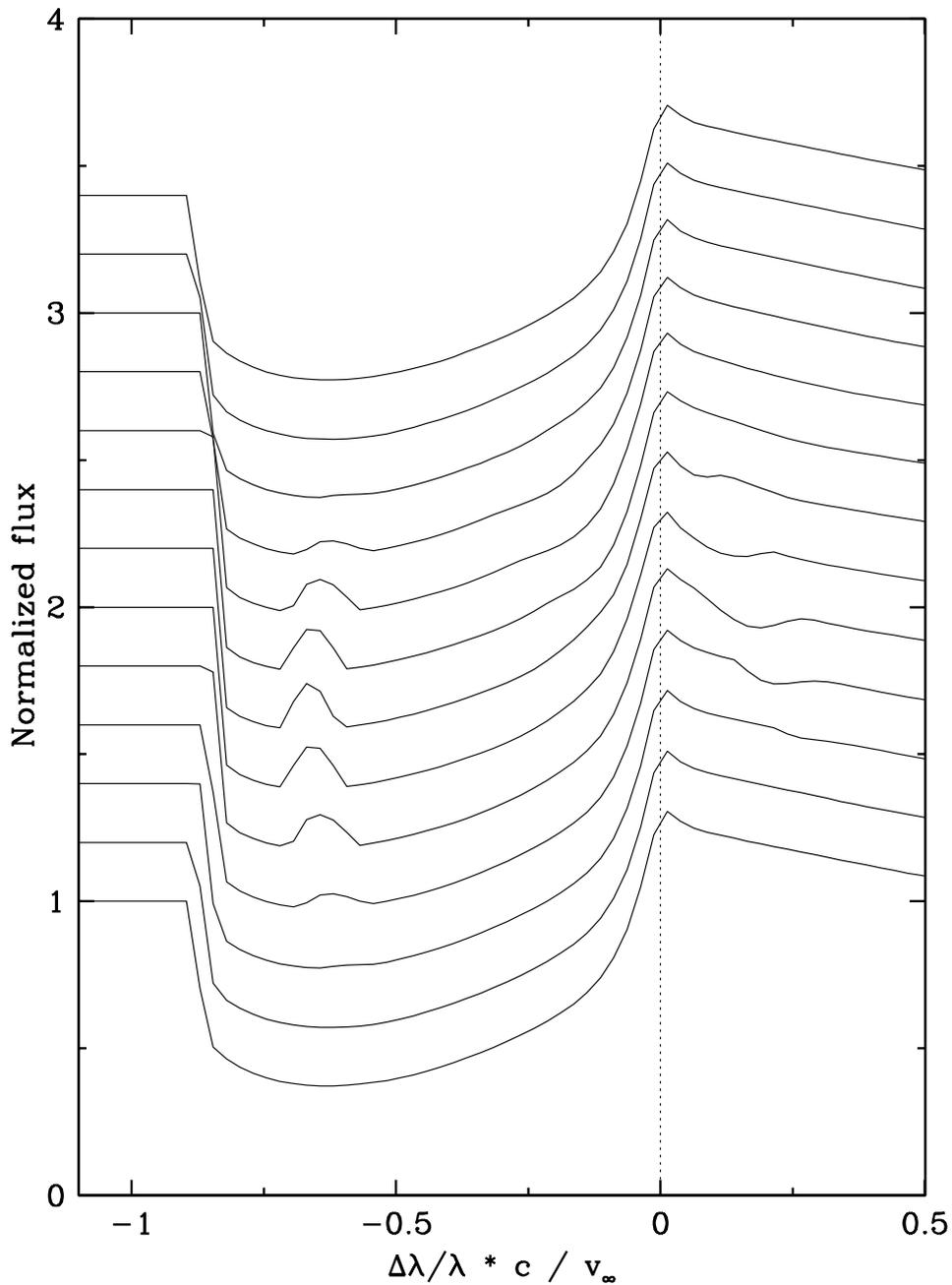}  
\caption{Line profiles computed with {\sc Wind3D} for the wind model shown in 
Fig.~\ref{fig blob}. 
The absorption in the P Cygni profile weakens because wind scattering in front of the 
stellar disk decreases when a local opacity enhancement (a spherical clump model) tangentially 
crosses the line of sight and partly obscures the star. The blob diminishes the wind scattering 
volume in the cylinder towards the observer where the absorption in the expanding wind 
locally diminishes ({\em see text}).
}
\label{fig spectrum blob}
\end{figure}

Further 3D transfer tests with parameterized spherical wind perturbations 
(`blobs' or `clumps') of denser gas that move in front of the star produce 
extra and less absorption at different wavelengths in the wind profile. 
In Fig.~\ref{fig blob} the opacity in 
the clump has been increased by an order of magnitude compared to the ambient wind opacity. 
The clump passes in front of the stellar disk and partly obscures it. 
It moves {\em perpendicular} ({\it tangentially drawn arrows}) to the surrounding 
radially expanding wind ({\it outer arrows}) which enhances the line absorption 
around the rest velocity (small dips in the line emission lobe). The dynamic spectrum 
in Fig.~\ref{fig spectrum blob} shows how the absorption portion of the P Cygni 
profile 
becomes weaker
when the local opacity enhancement in the blob crosses the observer's 
line of sight. However, the amount of photon scattering 
between the star and the observer also decreases because the blob removes a 
region in the wind where resonant scattering would occur at the velocity of 
the surrounding wind (as the blob moves perpendicular to the 
radial wind expansion in the model). Around $\sim$70~\% of $v_{\infty}$ the 
wind scattering therefore diminishes, yielding somewhat larger line fluxes in 
the absorption portion of the P Cygni profile (small bumps in the absorption trough). 
The clump diminishes the amount of line absorption in the line profile 
of Fig.~\ref{fig spectrum blob}, but also the line re-emission.
For a given line of sight the decrease of absorption 
due to the blob does not have to equal the decrease of emission. 
When the clump is located in front of the star the``loss of absorption exceeds the loss of 
emission since absorption of photons is re-emitted into 4 $\pi$.
Other calculations with {\sc Wind3D} show how the line absorption decreases further with 
larger clump opacities and larger clump sizes. The distance of the clump to the star determines 
the precise velocity position of the flux bump since the wind accelerates radially. 
In this example the velocity position of the small flux bump is (almost) invariable because 
the smooth wind velocity profile is spherically symmetric and the clump moves at 
the same distance from the star through the wind for an observer in the plane 
of the equator.

It is not a priori clear if a fully self-consistent line source function 
calculation is required in our modeling.
We are, after all, 
not concerned with detailed changes in the relative depth of DACs 
(our best fit procedure is based on the DAC shape and FWHM evolution),  
and far blue-shifted DAC absorption chiefly results from line opacity in the 
wind volume in front of the stellar disk. 
When the volume in this cylinder is relatively small compared to the 
total wind volume its emission can safely be neglected with respect to 
the contributions from the emission lobes. Since the height of our 
hydrodynamic wind models is only 1 $R_{*}$ around the plane of the equator, 
having almost {\em strictly radially} expanding wind structures, there is no far 
blue-shifted emission emerging from the emission lobes that can significantly 
alter the small DAC absorption. It is however not correct to assume that an 
iteration of the source function in 3D radiative transfer calculations can be neglected 
for detailed modeling of wind features observed in any type of spectral line.  
A stronger underlying P Cygni profile would be influenced more
by the emission, which could start to ``erase" the DAC at lower
velocities. Also, when
the absorption features occur at much smaller velocities (around stellar 
rest say), and the hydrodynamic structures yield far blue-shifted emission, 
the detailed source function of the structured wind cannot be neglected. 
This is for instance the case for the double-peaked emission lines formed in 
Be-star disk winds, or for the wine-bottle type H$\alpha$ emission profiles 
modeled by \citet{Hummel92}. The study of the physical properties of 
large-scale structures in winds of massive hot stars is not limited 
to UV resonance lines only. The H$\alpha$ profile of HD~64760 \citep{Kaufer+al06}
reveals symmetrically blue- and red-shifted emission humps with 2.4 d modulations 
that can result from variations at the base of the stellar wind. It remains 
to be studied how 3D radiative transfer modeling of H$\alpha$ can be combined with 
hydrodynamic models of rotational modulations observed in the Si~{\sc iv} line 
(but not addressed in this paper). It is clear however that to compare 
detailed normalized fluxes of both lines
a self-consistent calculation 
of the H$\alpha$ and Si~{\sc iv} line source functions in the structured 
wind model is required.

\subsection{Hydrodynamic Wind Models}\label{section hydro models}

Our approach of computing hydrodynamic CIR models is very similar to that of 
\citet{Cranmer+Owocki96}. The main differences are that we utilize the
{\sc Zeus3D}\footnote{{\tt http://www.astro.princeton.edu/$\sim$jstone/zeus.htm}}
\citep{Stone+Norman92} code rather than the VH-1 code, and that we 
introduce a spot velocity that does not have to equal the stellar rotational
velocity.

We first outline the procedure and summarize the assumptions that we share with
Cranmer \& Owocki. The time-dependent
3D equations of hydrodynamics are solved, limited to the equatorial 
plane of the star. Full 3D hydrodynamic calculations of CIRs
have been presented
by \citet{Dessart04} and these suggest that 3D effects are small
for spots that are symmetric around the equator. 
We opted to limit our calculations to the equatorial plane
as the gain in computing time allows us to more fully explore the parameter space.

Our models include the rotation in the equatorial plane of the star, the gravity acceleration (corrected for electron scattering),
and the line force due to radiative driving by line scattering. For evaluating the line force we 
apply the local Sobolev approximation and the CAK parameterization using 
a finite-disk correction. Note that this approach neglects the diffuse 
radiation field, which also suppresses the line-driven instability.
It allows us to investigate the effect of the 
large-scale CIR wind structures rather than small-scale
structures due to instability.
For the Sobolev approximation we apply the absolute value of the 
radial velocity gradient \citep{Rybicki+Hummer78}, while we neglect any
multiple resonance points due to the possibly non-monotonic wind velocity.
Although HD~64760 is a rapidly rotating supergiant 
we neglect possible effects of gravity darkening and rotational oblateness
on the star and its wind. Limb darkening is also neglected.
We include only the radial component of the radiative force in our
model. 

The neglect of non-radial forces, gravity darkening and oblateness
could be relevant for the mass loss rate in the equatorial
plane. \citet{Owocki+al96} have shown how material can flow poleward
if these effects are included (contrary to what might be expected from
the Wind-Compressed Disk model -- \citet{Bjorkman+Cassinelli93}).
The mass-loss in the equator could therefore be smaller than the
published value.

In our hydrodynamic model, one or more local radiation force 
enhancements (`spots') can be introduced at the base of the stellar 
wind. 
Each spot is determined by its position $\phi_0$ at time $t=0$ on the stellar 
surface, spot rotation velocity $v_{\rm sp}$, spot strength $A_{\rm sp}$ (or brightness), 
and opening angle $\Phi_{\rm sp}$ (the diameter of the spot). 
The spot is assumed to be 
circular symmetric
and centered on the equator. Only
the radial component of the additional line force due to this spot
is included in the model. The additional force is responsible
for the structure in the wind.
The line force takes into account the full extent of the spot (i.e. 
not only the part on the equator but also those parts above and below it).
The force is calculated
analytically for any point directly above the spot center
(including limb darkening in the spot).
For other positions it is assumed that a Gaussian
function (of azimuthal angle) relates the line force there to that
directly above the spot center; this assumption is made to reduce
computation times.
For detailed equations we refer to 
\citet{Cranmer+Owocki96}.

Our hydrodynamic modeling also includes radiative cooling in the energy conservation 
equation. Radiative heating has not been implemented, but instead 
a floor temperature is adopted to prevent the wind material from 
cooling down to below 80\% of the stellar effective temperature. 
However, in the models presented in this paper this cooling turned out
to be unimportant and all models are isothermal at 0.8 $T_{\rm eff}$.
We find in our exploratory calculations that this is no longer true for
very bright spots (brighter than is required for modeling HD~64760). 
As an example, for $\Phi_{\rm sp}$ = 20\degr~and $A_{\rm sp} > 1.2$, 
the wind is no longer isothermal due to shock heating. The 
boundary value for $A_{\rm sp}$ where shocks become important is
dependent of the spot angle and velocity. We find however
that the wind is isothermal for all models with $A_{\rm sp} \le 1.0$.

The main difference in our modeling with \citet{Cranmer+Owocki96} is 
that we allow the (angular) velocity of the spot rotation to differ 
from the velocity of the rotating stellar surface 
($v_{\rm sp}$ $\neq$ $v_{\rm rot}$). This is based on two arguments.
Firstly, \citet{Kaufer+al06} have shown
that a beat pattern of non-radial oscillations (in HD~64760) can produce 
local surface spots required for modeling the large-scale wind structures. 
The beat patterns, however, are not locked onto the stellar surface, 
but can rotate with a velocity different from the stellar rotational velocity.
Secondly, the observed recurrence time scale of the DAC 
compared to the 
estimated
rotation period
(Sect.~\ref{dacfit}) demonstrates that spot velocities 
considerably different from the surface velocity are required to match the
observations. By abandoning the usual assumption that 
$v_{\rm sp}$ equals $v_{\rm rot}$ a much wider range 
in the dynamic behavior of the large-scale wind structures emerges from the models.

We have thoroughly tested the new hydrodynamic code and can successfully 
reproduce the results of \citet{Cranmer+Owocki96}.
We tested that the results of the code are not sensitive to
details, such as the number of points in the grid, the extent
of the grid, the Courant number and the assumed floor temperature.
The results are slightly sensitive to the order of the advection
scheme that is used (by default we use a second-order scheme),
but the effect is not large
enough, however, to influence the conclusions of this paper.

The stellar and wind parameters of HD~64760 were adopted from \citet{Kaufer+al06}
and are listed in Table~\ref{table parameters}. 
Note that these parameters are in acceptable agreement with the more
recent ones derived by Lefever, Puls, \& Aerts (2007).
We determine the CAK $k$ and $\alpha$
parameters to obtain the observed mass-loss rate and the terminal wind 
velocity
(for simplicity, we assumed the CAK parameter $\delta=0$).
We apply a polar grid in the hydrodynamic calculations. 800 grid points 
sample the radial direction covering the wind range 1--30 $R_*$.
The grid-step increases by a fixed ratio from the inside to the outside
of the radius grid.
For the angular part 900 spatial angle points are uniformly distributed 
over the full $2\pi$ range. 
We set the Courant number to 0.5.
The initial conditions of the model start with an angle-independent smooth
wind flow having a $\beta$-velocity law with the observed $v_{\infty}$. 
The density structure of the smooth wind is derived from the conservation 
of mass and the observed mass loss rate. During the initial part of our 
hydrodynamic calculations the smooth wind settles into a stable steady-state outflow 
that becomes time-independent. Next one or more local spots are turned on at 
the stellar surface, yielding large-scale asymmetric structures in the extended stellar 
wind. We then proceed until the models assume a stationary state with a 
steadily expanding structured rotating wind. 

The resulting equatorial density and velocity structure are then
introduced into the {\sc Wind3D} code. This is done by copying
the structure in all planes parallel to the equatorial one, within
$\pm 0.5 R_{\rm *}$ around the equatorial plane.
Outside this region, the density and velocity have their smooth
wind value (as calculated by {\sc Zeus3D}).

\section{Observations}\label{obs}

We investigate the detailed DAC properties in the high-resolution time series 
of IUE spectra of the B-type supergiant HD~64760 (B0.5 Ib), observed during the 1995 MEGA campaign
\citep{Massa+al95a}. 
Detailed analyses of the 1993 and 1995 IUE MEGA campaign data of this star have been presented
in the literature. The 1993 data are thoroughly discussed in \citet{Massa+al95b}, while
\citet{Prinja+al95} and \citet{Howarth+al98} present time-series analyses of 
both data sets. These studies however provide periods for the rotationally modulated 
wind variations rather than for the slower evolving DAC structures we model in this paper. 
\citet{Fullerton97} presented a complete Fourier analysis of these data comparing 
detailed time-series of many lines individually, however also primarily focusing on the 
modulation periods. Only in a more recent study did \citet{Prinja+al02} point out 
that the recurrence time scale for the (slowly migrating high-velocity) DACs is not 
constrained. We therefore re-analyzed the 1995 MEGA campaign data with the goal 
of determining an accurate recurrence time for the DACs of HD~64760.   

The 1995 data sets are extracted from the IUE archive maintained at the Goddard Space Flight Center.
The IUE SWP spectra cover the wavelength range from 1150 to 1975~\AA\, with a nominal spectral resolution 
$R$ $\simeq$ 10,000.  The spectra were created by staff at GSFC using the IUEDAC software to combine 
most of the echelle 
orders to form a single continuous spectrum, which was placed on a uniformly 
spaced wavelength grid with 0.05 \AA\, spacing. The wavelength scale calibration 
was improved by centering the spectra on selected interstellar lines using an echelle order-dependent 
correction (see also \citet{Prinja+al02}). 
The mean signal-to-noise ratio in the continuum is typically 
$\sim$20 to 30 for both data sets. 

\begin{figure}  
\plotone{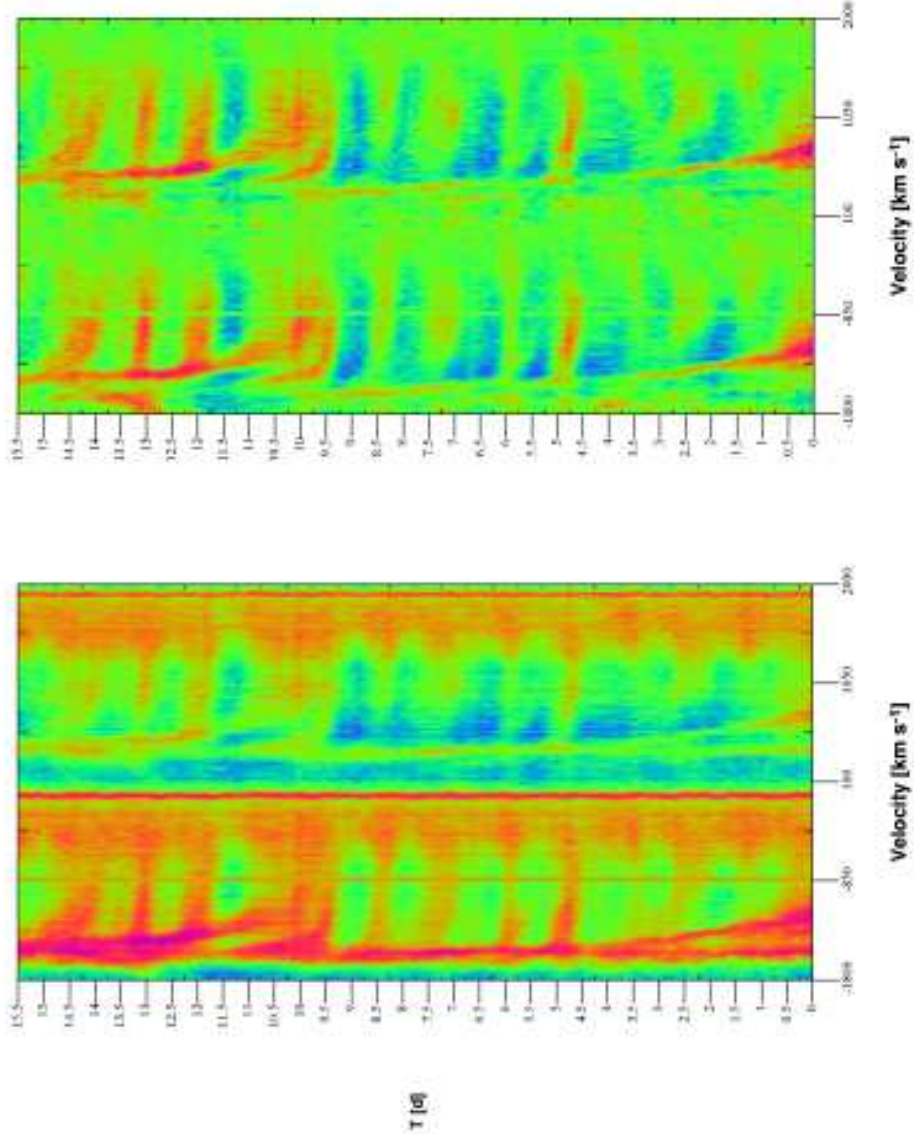}  
\caption{{\bf Left-hand panel:} Time sequence of the IUE normalized flux spectrum of HD~64760 
(time runs upward) of the Si {\sc iv} $\lambda$1400 resonance doublet in velocity scale 
centered around the short-wavelength line. Dark and bright shades indicate low and high 
flux levels. {\bf Right-hand panel:} The flux difference spectrum shows DACs drifting 
blueward from velocities exceeding $\sim$ $-$1000 $\rm km\,s^{-1}$ to $-$1600 $\rm km\,s^{-1}$.
The depth and width of the DACs decrease over time.       
They assume the shape of a `slanted triangle' over a period of $\sim$3~d. The top of the triangle  
extends further into a `tube-like' narrow absorption feature that remains visible over the next 7~d 
and drifts asymptotically to a maximum velocity.
}
\label{fig 64760 spectra}
\end{figure}

We combine a time series of 145 SWP spectra observed between 
1995 January 13 at UT 12:57:35 (JD 2449731.03999) and January 29 at 01:02:18 (JD 2449746.54326).
The spectra are integrated with the large aperture during 60 s, and observed about every 3 hours 
over this period of 15.5~d. The left-hand panel of 
Fig.~\ref{fig 64760 spectra} shows the time-sequence of the 
flux spectrum in velocity scale centered around the 
short-wavelength line of the Si {\sc iv} $\lambda$1400 resonance doublet. 
The dynamic spectrum is plotted with grey-scales for which dark and bright shades 
indicate low and high flux levels, respectively. The dark and bright shades correspond to
red (or violet) and blue colors in the online image versions. The dynamic spectra 
in Fig.~\ref{fig 64760 spectra} are linearly interpolated over time to provide a 
uniform time-sampling. The dynamic spectrum reveals many subtle spectral features 
in the absorption portion of the P Cygni profile of both Si {\sc iv} lines. 

The straight vertical lines around 0 and $+$1900 
$\rm km\,s^{-1}$ are due to interstellar absorption of Si {\sc iv} toward HD~64760. 
We plot the velocity scale in the stellar rest frame using 
a radial velocity of 41 $\rm km\,s^{-1}$ (CDS Simbad), and a small heliocentric correction 
($\sim$5.43 $\rm km\,s^{-1}$ from GSFC data header files). 
The flux minimum in the interstellar lines is constant 
within a fraction of a percent of the local continuum flux which indicates that 
the absolute flux calibration of the sequence is reliable.      
Note that the straight vertical line around $-$800 $\rm km\,s^{-1}$ in the left-hand 
panel of Fig.~\ref{fig 64760 spectra} 
is due to a detector reseau mark. The zero flux values for these bins in the spectra
have been set to a fixed value to prevent a compression of the
grey-scale range in the dynamic spectrum image.                  
 
We investigate the Si {\sc iv} lines of HD~64760 because they exhibit a variety of 
remarkable spectral features that signal formation in dynamic wind structures. Most striking
are the slowly migrating high-velocity DACs observed between $T$=0 and $\sim$5
d (further `lower DAC'), and between $T$=10 and 15~d (`upper DAC') 
in Fig.~\ref{fig 64760 spectra}. Both Si {\sc iv} resonance doublet lines are 
sufficiently well separated. The blue edge velocity of the 
long wavelength P Cygni line profile does not overlap with the flux evolution of the DAC observed 
in the short-wavelength line. The S/N ratios in the Si {\sc iv} wavelength region are sufficiently 
large to permit image processing techniques to the dynamic flux spectrum to investigate
the detailed DAC properties without the application of any compromising spectral smoothing 
operations. In the right-hand panel of Fig.~\ref{fig 64760 spectra} 
we subtract the mean flux per wavelength 
bin from the absolute flux image in the left-hand panel. 
This operation removes the overall 
shape of the underlying P Cygni profiles of both doublet lines. The strong absorption  
feature observed around the blue edge velocity of $\sim$ $-$1600 $\rm km\,s^{-1}$ 
is therefore removed in the right-hand panel of Fig.~\ref{fig 64760 spectra}. 

The flux difference image shows that the DACs in 
both doublet lines assume comparable depths and also reveal an almost identical 
flux evolution over time (see also the flux difference spectrum in Fig.~9 of \citet{Fullerton97}). 
They drift bluewards from velocities exceeding 
$\sim$ $-$1000 $\rm km\,s^{-1}$ to $-$1600 $\rm km\,s^{-1}$.    
The depth and width of the DACs decrease over time while drifting bluewards.       
The DACs are rather broad and intense when they appear in the P Cygni absorption line 
portion and gradually weaken and narrow while shifting bluewards. The base of the 
DAC evolution in HD~64760 reveals the shape of a slanted triangle over a period 
of $\sim$3~d. The top of the triangle further extends into a `tube-like' narrow absorption 
feature that can be traced over the next 7~d and drifts asymptotically to a maximum 
velocity that slowly approaches the blue edge velocity of the P Cygni line profile.    
This is considerably clearer in the difference spectrum than
in the original one.
The flux difference spectrum clearly reveals that the lower DAC extends over time to at least 
$T$ $\simeq$ 10~d.

\begin{figure}  
\plotone{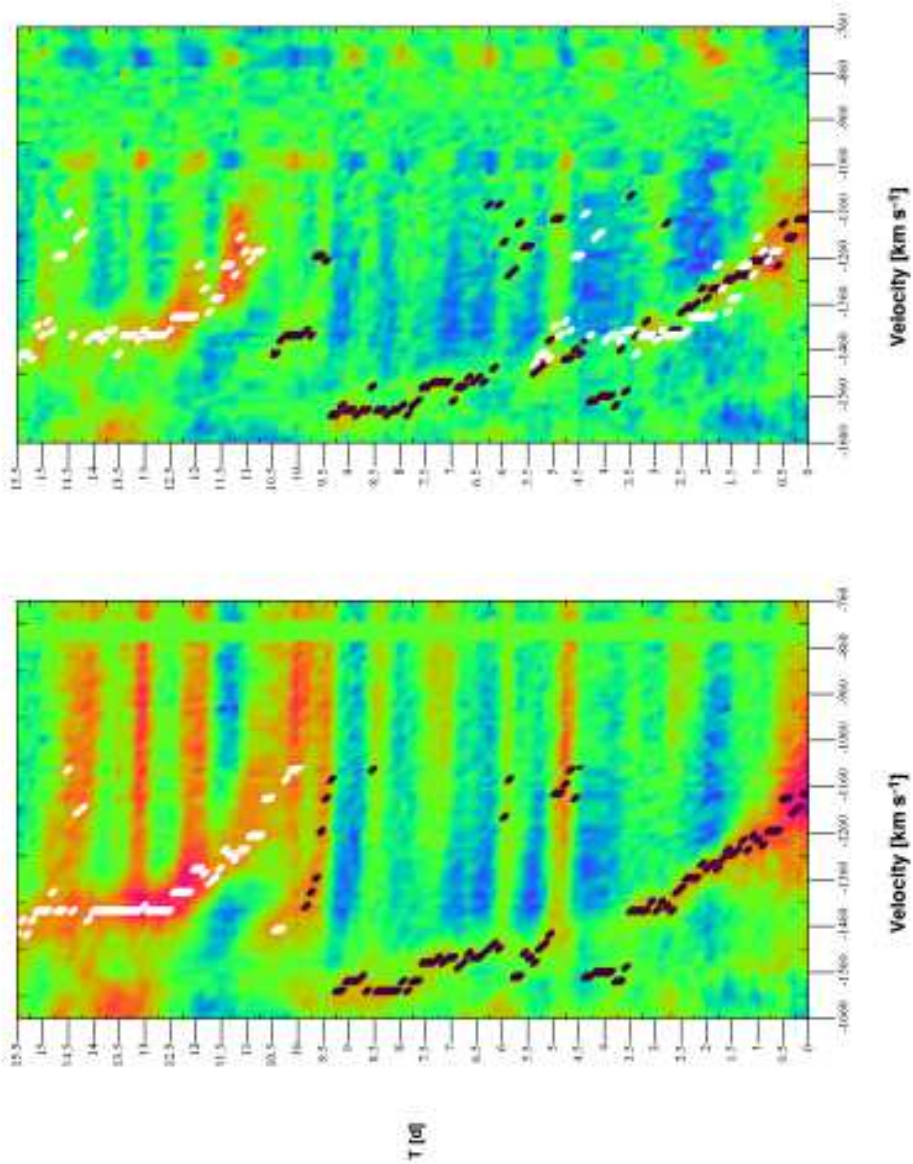}  
\caption{{\bf Left-hand panel:} HD~64760 flux difference spectrum of the short-wavelength
component of the Si~{\sc iv} doublet. 
The DAC minima of the lower DAC are indicated with black dots, those of the upper DAC
with white dots. Around $T\simeq$ 10~d ({\em white dots}) the DACs are distorted by
horizontal rotational modulations, complicating a reliable determination of the DAC recurrence time. 
{\bf Right-hand panel: } The flux contributions of the horizontal modulations are canceled out by mirroring 
and subtracting
the left-hand image. Black dots mark the flux minima in the lower DAC, white dots the upper DAC.
They are also shown shifted downwards for a best match. 
The DAC recurrence time of 10.3$\pm$0.5~d is more accurately constrained from this image rather
than from the left-hand image.
}
\label{fig DAC minima}
\end{figure}

We next constrain the recurrence time-scale for the DAC in HD~64760. For our DAC modeling purposes 
in Sect.~\ref{dacfit} it is important to establish an accurate period over which the DAC
recurs.
In the right-hand panel of Fig.~\ref{fig 64760 spectra} 
we observe besides the DACs several 
individual spectral features that cover a large velocity range. They reveal a
rather `bow-shaped' morphology, extending bluewards and redwards at the 
same time. The `horizontal bow' shape of these features clearly differs from   
the more `triangular' shape of the DACs. These `modulations' exhibit a possible period in HD~64760 
of $\sim$1.2~d \citep{Fullerton97,Prinja+al02}, although careful inspection of the flux difference image 
reveals that these shallow modulations alternate over time 
with comparable absorption features that extend even more horizontally 
(i.e. they are almost completely horizontal without the rounded bow-like shape). 
Clear examples of both types of
horizontal features are observed around 4.8~d and 6~d, and around 7.3~d and 8.5~d.        
Similarly shaped, but stronger modulations, occur around 12~d and 13.2~d when they distort 
the tube-like narrow absorption in the DAC at larger velocities. 
We observe that the flux differences in the bow-shaped modulations are distributed almost symmetrically 
around a velocity axis that is located $\sim$ $-$930 $\rm km\,s^{-1}$. For example, the 
fluxes in the modulation observed around 4.8~d decrease nearly symmetrically blueward and redward 
of this velocity value. We use this near-symmetry property of the bow-shaped features to remove the disturbing 
flux contributions from the modulations to the detailed DAC evolution, and to increase the flux 
contrast of the DAC. In the right-hand panel of 
Fig.~\ref{fig DAC minima} we subtract 
the mirror image from the left-hand panel. The velocity position of the 
mirror axis at $\sim$ $-$930 $\rm km\,s^{-1}$ is determined by minimizing the flux contributions in the horizontal 
features observed between 11.5~d and 15~d that distort the flux values across the upper DAC. 

Our mirroring flux difference procedure effectively cancels out the flux contributions 
from the modulations, while the fluxes inside the DAC remain unaffected (since the DAC 
is not observed symmetrically around the selected mirror axis).
%We can determine the recurrence time of the DAC more accurately from the
%fluxes corrected in this way because our automatic procedure to find
%the flux minimum at each time will in most cases find the minimum of
%the DAC, rather than that of a perturbing feature. 
We next apply an automatic procedure to find the flux minimum for each
separate spectrum. Due to our mirroring correction, this procedure will
in most cases find the minimum of the DAC rather than that of a 
perturbing `modulation' feature.
The minima thus found are indicated
in black (lower DAC) and white (upper DAC). In the right-hand panel
of Fig.~\ref{fig DAC minima} we
shift the upper DAC minima downwards and compute 
the best match with the shape of the lower DAC minima. 
We apply a simple least-squares minimization technique to the velocity positions of 
the shifted upper and the lower flux minima and find 
the best match for $\Delta T = 10.3$~d.
We hence determine a recurrence time of 10.3~d for the DAC in 
Si~{\sc iv} $\lambda$1395 of HD~64760. 
We estimate an error bar of $\pm 0.5$~d on this result.
It is clear from Fig.~\ref{fig DAC minima} that the upper DAC is
not a repeat of the lower DAC. 
We show in Sect. 5 that a hydrodynamic wind model with two unequal CIRs 
provides the best fit to the shape and morphology of the lower and upper DAC.

It is of note that the detailed shape of the high-velocity DACs observed 
in HD~64760 strongly resembles the shape of DACs observed in the Si~{\sc iv} lines
of $\xi$ Per (O7.5 III). The peculiar slanted triangular shape of the DAC base, 
which extends into a tube-like absorption feature, is also observed in a sequence of 
IUE spectra of 1991 October \citep{Kaper+al99}. Interestingly these data reveal
that the DACs in $\xi$ Per can extend to velocities considerably below the 
values observed for the DAC in HD~64760. In HD~64760 DAC velocities are observed 
only blueward of $\sim$ $-$1000 $\rm km\,s^{-1}$, while in $\xi$~Per they reach 
almost zero velocity, or down to the very base of the stellar wind. The foot of 
the DAC base in $\xi$~Per is nearly horizontal (e.g. the boundary where the DAC 
first appears is very sharp) signaling that the dynamic structures that produce 
the DACs can extend geometrically very far through the wind when they start to 
rotate into the observer's line of sight (from almost the base of the wind to  
$\sim$ $v_{\infty}$).

In the following Sections we compute hydrodynamic models that correctly fit the 
detailed shape and morphology of the DACs in HD~64760. We find that the peculiar 
DAC shape cannot be computed with (simple) parameterized models of the wind 
density in Sect.~\ref{section parameterized models}. A detailed hydrodynamic 
model of $\xi$~Per will be presented in a future paper.

\section{Hydrodynamic Wind Modeling}\label{grids}

\subsection{Model Example}

\begin{figure}  
\plotone{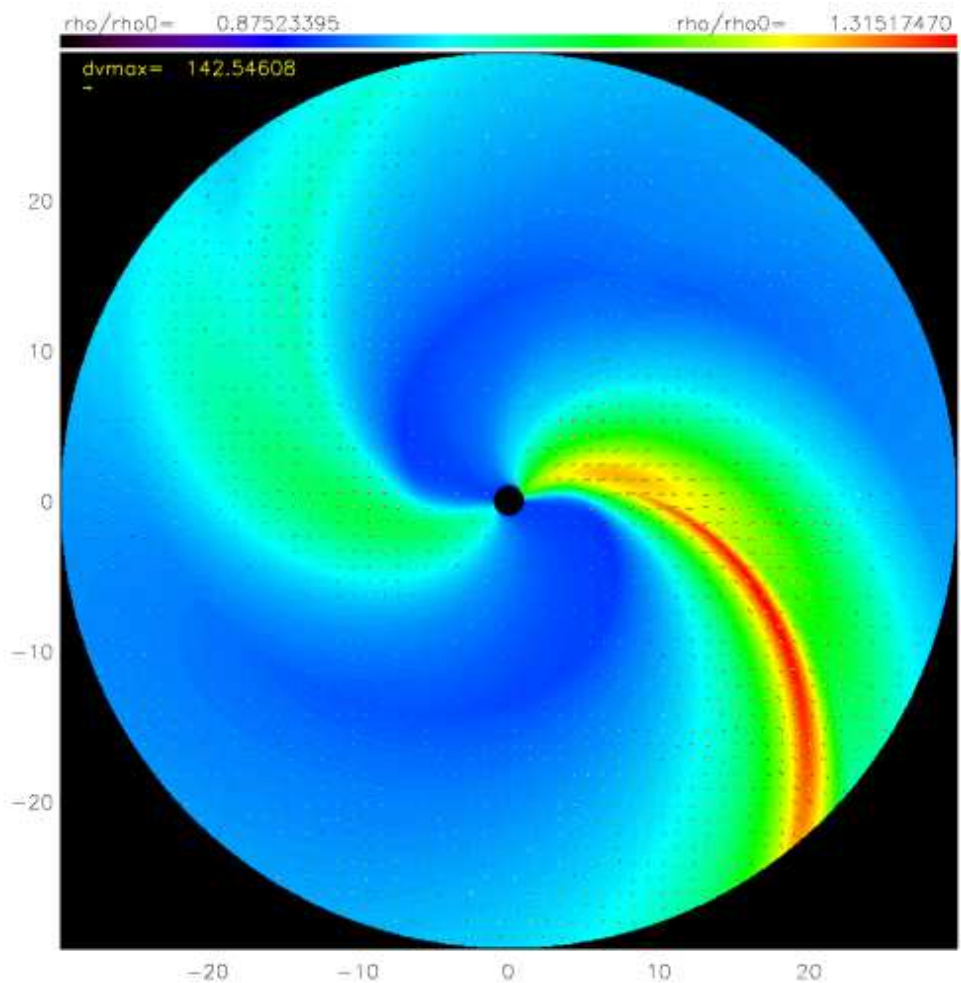}  
\caption{Hydrodynamic wind model with large-scale density- and velocity-structures computed with {\sc Zeus3D}
for HD~64760 required to best fit the DACs in Si~{\sc iv} $\lambda$1395. The bright spots at the equator
are 20\% and 8\% brighter than the stellar surface, 
having opening angle diameters of 20\degr~and 30\degr, respectively,
and rotate 5 times slower 
than the surface. The spots produce wind structures with density enhancements in CIRs compared 
to the smooth wind density. The minimum density contrast is 0.87 ({\em dark shades of grey}), 
increasing to a maximum value of 1.32 ({\em bright regions}). The size of the over-plotted arrows indicates 
the magnitude of the velocity deceleration with respect to the smooth unperturbed wind. The maximum 
velocity difference in the CIRs compared to the smooth wind
does not exceed $\sim$~140 $\rm km\,s^{-1}$. The bright spots cause density waves in the equatorial 
rotating wind.
[{\em This figure is available as an animation in the electronic
 version of the Journal}
]
}  
\label{fig best fit}
\end{figure}

Figure~\ref{fig best fit} 
shows the density contrast for the hydrodynamic 
model\footnote{Animations of a number of hydrodynamic models are available at
               {\tt http://www.astro.oma.be/HOTSTAR/CIR/CIR.html}
              }
of HD~64760 
with $v_{\rm sp} = v_{\rm rot}/5$, having two spots 
that are 20\% and 8\% brighter than the stellar surface ($A_{\rm sp}$=0.2 and 0.08),
with spot angle diameters of $\Phi_{\rm sp}$=20\degr~ and 30\degr, respectively.
The wind velocity deceleration is typically largest inside the large-scale dynamic structures 
where the density contrast is also largest. This results from the conservation of 
momentum in the hydrodynamics equations. For an observer in the laboratory frame, 
wind particles expelled at the stellar surface follow 
almost straight radial paths outward. When the wind particles cross the spiraling structure radially 
they temporarily decelerate with respect to the smoothly accelerating wind, thereby enhancing 
the wind density locally. 
The extra mass from the bright spots 
induces stable density waves through which the surrounding fast wind streams. 
The density waves rotate in the plane of the equator with a period set by that of the spots. 
The tangential wind velocity components are small compared to the radial velocity 
components. They become relatively largest close to the stellar surface where the radial wind 
velocity is smallest. In this model the local deceleration of the flow 
across the density spirals does not exceed $\sim 140$ $\rm km\,s^{-1}$ and decreases outward 
along the large-scale wind structures, becoming vanishingly small at radial distances beyond 
30 $R_{*}$ where the density approaches that of the smooth wind expanding at $v_{\infty}$.

\subsection{Model Grid}\label{section model grid}

To understand the effects of the different input parameters on the hydrodynamics, 
we compute a large grid of models with the {\sc Zeus3D} code.  We first compute the 
hydrodynamic model for the stellar parameters of HD~64760 and its smooth wind properties
(Table 1). Next we introduce in the model a 
{\em single}
spot at the equator. We 
vary the spot parameters of brightness ($A_{\rm sp}$), spot angle ($\Phi_{\rm sp}$), and 
spot velocity ($v_{\rm sp}$) that provide different models for the large-scale structured wind. 
These models then serve as input to the {\sc Wind3D} code, which calculates the
resulting dynamic spectra. The underlying smooth-wind profile is subtracted from
all calculated profiles to allow a comparison to the observed flux difference
spectra. 

\begin{figure}  
\plotone{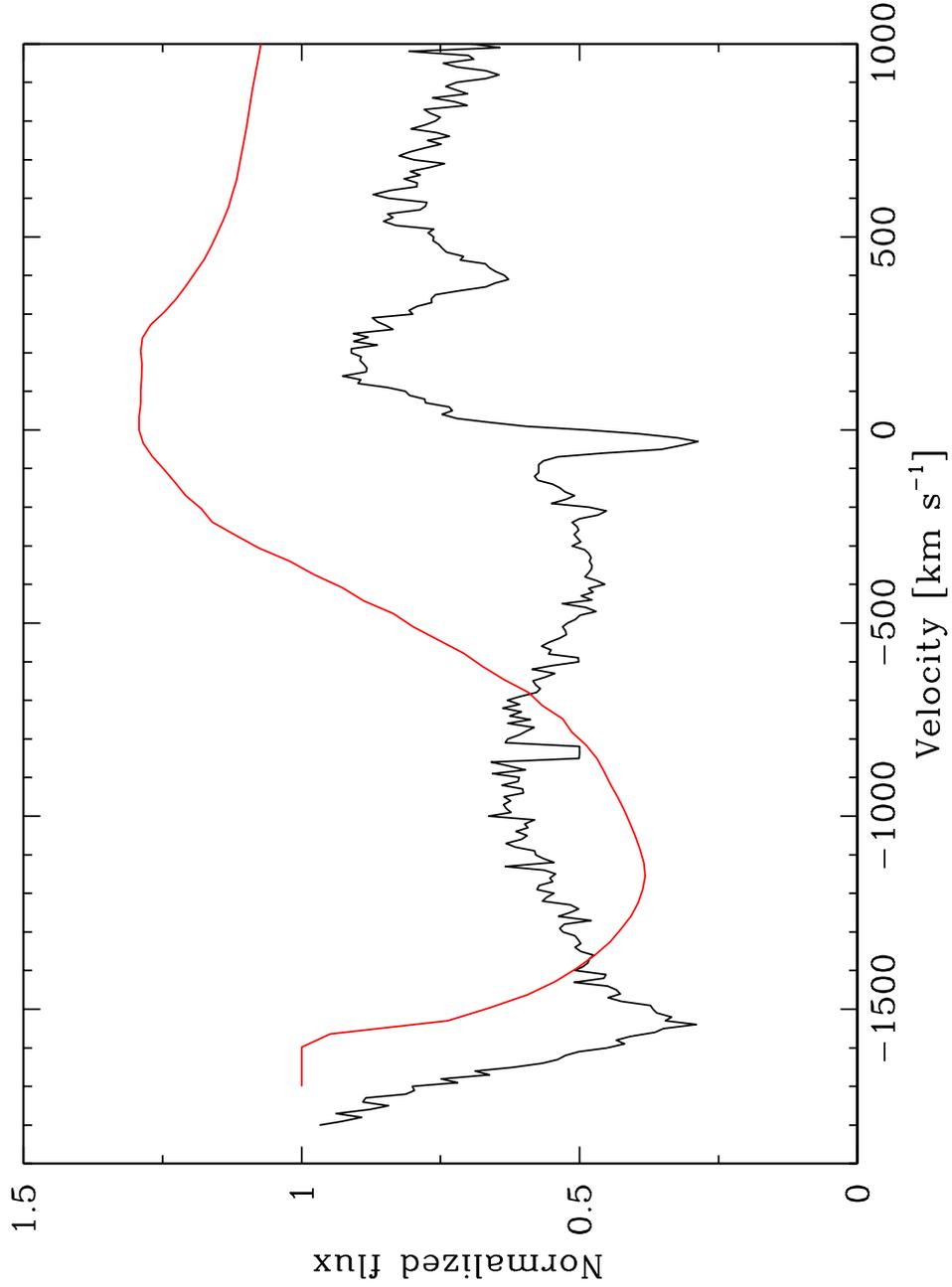}  
\caption{Mean normalized flux of Si~{\sc iv} $\lambda$1395 observed with IUE in HD~64760 during 15.5~d in 1995.
The average flux of the computed dynamic spectrum that fits the observed DAC shape and morphology in the 
line is over-plotted.}
\label{fig underlying profile}
\end{figure}

In our transfer calculations we compute the structured wind opacity in the Si~{\sc iv} 
resonance lines from the wind density contrast since  
$\overline{\chi^{l}}$/$\overline{\chi^{l}_{\rm smooth}}$ $\simeq$ $\rho$/$\rho_{\rm smooth}$ 
(as long as Si {\sc iv} is the dominant ion, which is the case
for HD~64760).
The smooth wind opacity $\overline{\chi^{l}_{\rm smooth}}$ in the lines is given by the
\citet{Groenewegen89} parameterization (see Appendix B).
We set the thermal broadening to a small value of 8 $\rm km\,s^{-1}$.  
When the opacity parameters are set to $T_{\rm tot}=1$, $\alpha_1=1$, $\alpha_2=1$, 
we find an acceptable ``fit" to the underlying (unsaturated) P Cygni profile of the 
UV Si~{\sc iv} lines. Figure~\ref{fig underlying profile} shows the average normalized flux 
profile of the Si~{\sc iv} $\lambda$1395 line of HD~64760 observed over a period 
of 15.5~d in 1995. 
It is of note that the Si~{\sc iv} line also contains other photospheric absorption 
lines (i.e. around $\sim$500 $\rm km\,s^{-1}$), together with an interstellar 
absorption line (around the stellar rest velocity), a detector reseau mark, and a 
steady strong absorption feature around $-$1500 $\rm km\,s^{-1}$, which cannot all 
be accounted for in our 3D hydrodynamic wind models. We emphasize however that 
our DAC modeling is based on a best fit procedure in the absorption portion of 
the unsaturated Si~{\sc iv} P Cygni line profile. The best fit procedure is therefore 
based on the observed and computed flux difference profiles, instead of the normalized 
flux profiles shown in Fig.~\ref{fig underlying profile}.

\subsection{Parameter Study}

In this Section we discuss the influence of the various parameters on the
CIR hydrodynamics, and how the shape and morphology of the DACs in HD~64760
changes. 
We concentrate on the spot parameters as we assume all other parameters
to be known. This parameter study
will be important in Sect.~\ref{dacfit}
where we determine the best fit to the observed DACs.
In doing so we concentrate on the detailed DAC shape (velocity position of the flux
minima with time) and morphology (DAC FWHM evolution over time).
We do not fit the detailed intensity variation of the DACs because 
that also depends on the opacity in wind structures above and below the 
equatorial plane we currently do not incorporate in our hydrodynamic models. 
We demonstrate however that detailed fits to the precise DAC depth are not 
needed primarily because the DAC shape and morphology are uniquely determined by 
the spot parameters.  

\subsubsection{Spot Strength $A_{\rm sp}$}

From the small grid of hydrodynamic models calculated by 
\citet{Cranmer+Owocki96}, it is already clear that
the effect of a larger spot brightness is
to increase the range of density contrast
(the ratio of density of the structured and the smooth wind $\rho$/$\rho_{0}$).
The density contrast maxima increase while the minima decrease.
Furthermore, the variations of the wind
velocity with respect to the smooth wind velocity also increase with 
larger $A_{
\rm sp}$.
Our more extensive grid confirms their results.
For an example of a spot with $v_{\rm sp} = v_{\rm rot}$
and $\Phi_{\rm sp}=20\degr$,
an increase of 
$A_{\rm sp}$ from 0.1, 0.3, 0.6, to 1.0 increases the maximum of $\rho$/$\rho_{0}$ 
from 1.5, 2.0, 5.0, to 13.7, respectively. 
The maximum velocity difference increases from 130 to 390 km\,s$^{-1}$. 
This trend is directly related to the local increase of the total mass-loss rate 
the spot causes by additional radiative wind driving
(see also Sect.~\ref{discuss}). At the stellar surface 
the spot injects extra material into the wind that collides with smooth-wind 
material, thereby locally increasing the density in the resulting large-scale wind structures. 
The larger the spot brightness, the more material is injected into the wind, and the 
stronger the dynamics of this collision.

In Fig.~\ref{fig Asp effect} we show the effect of $A_{\rm sp}$ on the
detailed DAC structure computed with {\sc Wind3D}.
We increase the spot intensity $A_{\rm sp}$ from 0.1 ({\em upper left panels}),
over 0.3 ({\em upper right panels}) and 0.6 ({\em lower left panels}),
to 1.0 ({\em lower right panels}). The spot angle $\Phi_{\rm sp}$=20\degr, while 
the spot velocity is set equal to the surface rotation velocity ($v_{\rm sp} = v_{\rm rot}$) in the four models. 
We plot both the hydrodynamic models and the dynamic spectra.
The hydrodynamic model shows the density contrast and velocity vectors with
respect to the smooth wind. The model rotates counter-clockwise over one period. 
The dynamic spectra show the rotation phase from 0.0 to 1.0. 
The rotation phase zero corresponds to the spectrum we compute for an observer in the plane of the equator 
viewing the rotating hydrodynamic model edge-on from the south side in these images. The flux difference 
profiles are shown between 0 and $-$1700 $\rm km\,s^{-1}$, which is the absorption portion of the P Cygni line profile.

\subsubsection{DAC Formation Region}

\begin{figure}  
\plotone{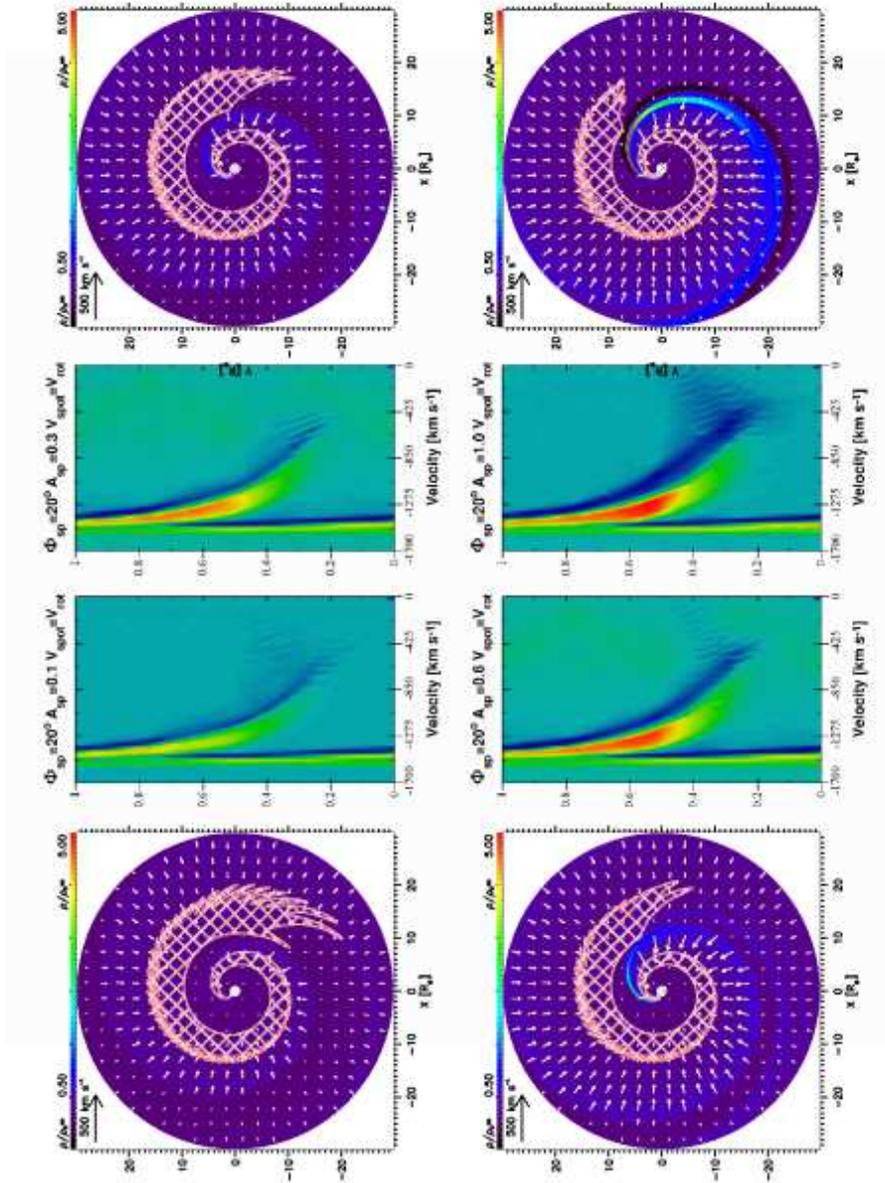}  
\caption{The spot intensity $A_{\rm sp}$ is increased from 0.1 ({\em upper left panels}), 
0.3 ({\em upper right panels}) and 0.6 ({\em lower left panels}), to 1.0 ({\em lower right panels}). 
The hydrodynamic models show the density contrast and velocity vectors with respect to the smooth wind.
The dynamic spectra show the rotation phase from 0.0 to 1.0 ({\em time runs upward}).
Rotation phase zero corresponds to the spectrum we compute for an observer in the plane of the equator
viewing the rotating hydrodynamic model edge-on from the south side in these images.
The formation regions of the DAC in the spectra are located behind the CIR in these 
hydrodynamic models ({\em hatched areas}). The increase of $A_{\rm sp}$ extends the DAC towards smaller velocities 
across the spectra. Hatched areas are those with high Sobolev optical depth ({\em see text}).}
\label{fig Asp effect}
\end{figure}

To study the formation region of the DAC, we introduce the ``relative"
Sobolev optical depth:
\begin{equation}
\frac{\tau}{\tau_0} = \frac{\displaystyle\frac{\rho}{|dv/dr|}}{\displaystyle\frac{\rho_0}{|dv/dr|_0}}
\label{eq Sobolev}
\end{equation}
where the subscript `0' refers to the smooth wind model, and we neglect the 
finite size of the stellar disk by considering only the central ray.
DACs are formed in wind regions with large $\tau/\tau_0$ values.
These regions are indicated in Figs.~\ref{fig Asp effect},
\ref{fig Phisp effect}, \& \ref{fig vsp effect} by hatched areas on the
hydrodynamic plots. It follows from Eq.~(\ref{eq Sobolev}) that both the
density and velocity gradients can play an important role for the formation 
of DACs.

As first pointed out by \citet{Cranmer+Owocki96}, the DACs are more
likely to be due to velocity plateaus than to the increased density 
in the CIR. The CIR causes a discontinuity in the velocity gradient
(``kink"). This kink moves upstream and therefore trails the density 
enhancement of the CIR. The velocity plateau is bounded on one side
by the kink and extends on the other side towards the CIR
(see, e.g., Fig.~\ref{fig Asp effect}). Because of the upstream
movement of the kink and velocity plateau, they are more warped
than the CIR. As the spot rotates, different parts of the velocity
plateau cross the line of sight at increasing wind velocities. If the DAC is
completely formed in the velocity plateau, this directly translates
to the DAC drifting toward larger velocities.
Because of the strong spiral
winding, the DAC velocity increases only slowly with time, which explains
the slow apparent acceleration of DACs 
\citep[see also][]{Hamann01}.

Our results confirm that DACs are mainly formed in the velocity plateau. 
As an example, consider
the upper right-hand panel of Fig.~\ref{fig vsp effect}. The dynamic spectrum
shows a DAC that persists over the whole cycle (starting at low
velocities around phase 0.25 and then narrowing with increasing phase until
it disappears around phase 1.3). The hydrodynamic plot, however,
shows a CIR that persist only over somewhat more than a quarter of the cycle.
The density enhancement in the CIR can therefore not be
responsible for the DAC.
Looking in more detail we find, e.g., that at phase 0 the DAC is present
around $-1500$ $\rm km\,s^{-1}$. On the hydrodynamic plot there is no 
density enhancement at all around phase 0 (south direction). 
The large Sobolev optical depth (hatched area) at this phase is due to 
the small value of the velocity gradient (i.e. a velocity plateau).

However, in certain cases it is the CIR density
enhancement that creates the DAC. We again refer to the example
of the upper right panel of Fig.~\ref{fig vsp effect}. A more detailed
study of this case shows that the
relative Sobolev optical depth ($\tau/\tau_0$) is dominated by the density
enhancement when the DAC is at low velocities (phases 0.25-0.5).
At larger velocities, the velocity plateau dominates in the formation of the DAC.

An important aspect of the hydrodynamic models is that an
increase of $A_{\rm sp}$ (all other model parameters being equal) enhances the
density contrast
$\rho$/$\rho_{0}$ inside the CIR {\em and} 
decreases
the corresponding radial wind
velocity gradient
behind the CIR (i.e. the wind regions between the stellar surface and the
CIR).
An increase of $A_{\rm sp}$ over the four models in Fig.~\ref{fig Asp effect}
causes the CIR with large $\rho/\rho_{0}$ to occur farther away from the
stellar surface and its overall curvature to diminish.
The wind region with large $\tau/\tau_0$ also shifts farther away from the denser
CIR.
In regions where $|\frac{dv}{dr}|$ is small the local outflow velocity close
to the stellar
surface considerably decreases because of the larger density inside the CIR.
For this reason strong DAC absorption in the dynamic spectrum extends further
towards smaller velocities
when the spot strength $A_{\rm sp}$ increases.

The radial approximation we introduced in Eq.~(\ref{eq Sobolev})
only provides an estimate of the DAC line formation depth.
It breaks down in wind
regions where non-radial
velocities are large compared to the overall radial wind expansion. 
We emphasize however that the DACs we compute in the dynamic spectra with {\sc
Wind3D} do not utilize the
radial approximation. Furthermore they also include the effect
of a (small) intrinsic Doppler broadening in the source function
and therefore go beyond the Sobolev approximation.
The line source function in the transfer is fully 3D
lambda iterated to
self-consistency with the radiative transfer equation to compute the detailed
DAC fluxes.

\subsubsection{Spot Opening Angle $\Phi_{\rm sp}$ }\label{spot opening angle}

\cite{Cranmer+Owocki96} conclude from their hydrodynamic models that the maximum density {\em decreases}
with increasing spot angle ($\Phi_{\rm sp}$) with the exception
of the smallest $\Phi_{\rm sp}$, where they presume to have undersampled
the wind structure. Our more extensive grid (also calculated at a higher
spatial resolution) shows a clearer picture.
For small spot angles, the range in density contrast and the velocity 
differences {\em increase} with increasing spot angle. 
This is consistent with the idea that
larger spot angles also increase the amount 
of mass injected into the surrounding stellar wind. 
Towards even larger spot angles, however, the larger density contrast 
and velocity differences begin to level off and to decrease
(consistent with the findings of Cranmer \& Owocki).
An example with $v_{\rm sp} = v_{\rm rot}$ and $A_{\rm sp}=0.5$ 
shows that the maximum $\rho$/$\rho_{0}$ varies from 1.2, over 3.8, down to 1.9
when $\Phi_{\rm sp}$ increases from 5\degr~over 20\degr~to 180\degr.
The decrease of $\rho$/$\rho_{0}$ with larger $\Phi_{\rm sp}$ results from the 
larger angular extent over which the extra mass injected by the spot
spreads out, which makes its collision with the ambient smooth wind 
material less efficient
(also noted by Cranmer \& Owocki).

\begin{figure}  
\plotone{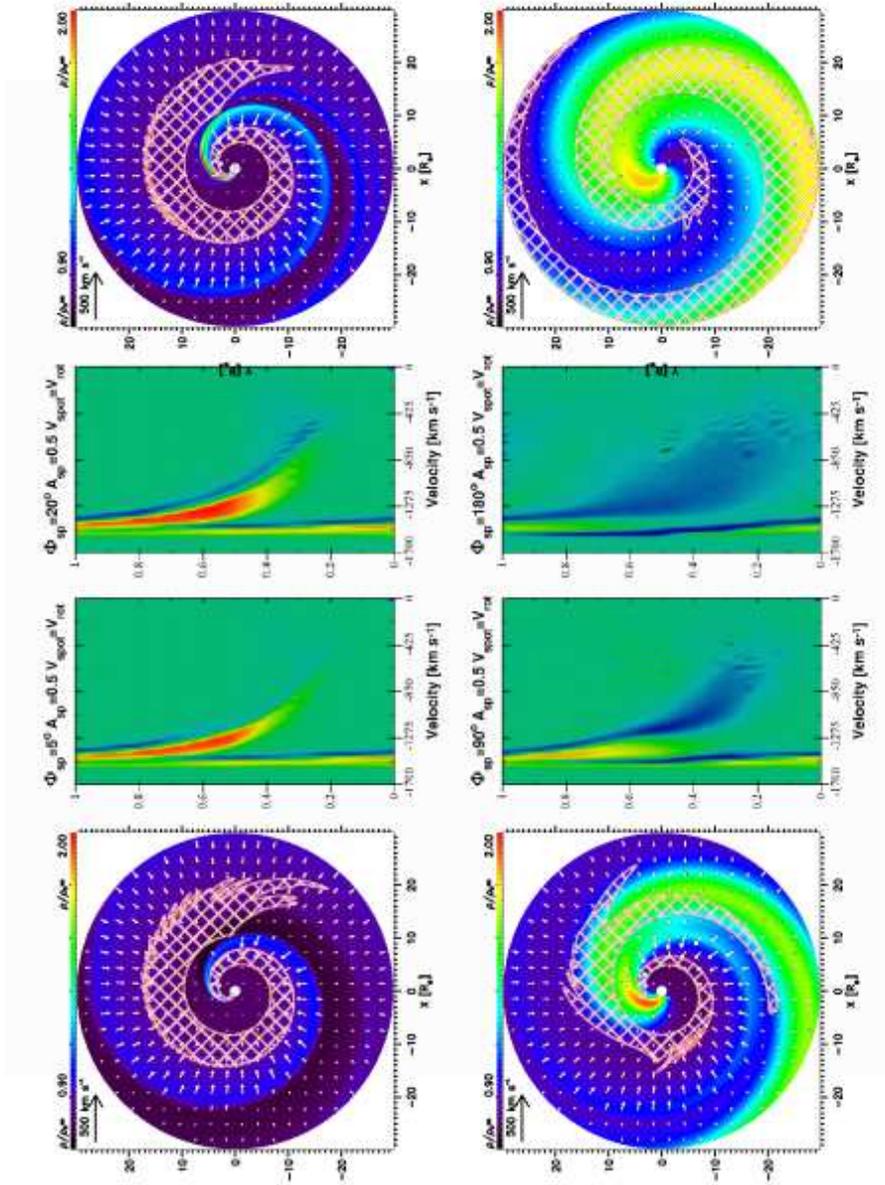}  
\caption{The spot opening angle $\Phi_{\rm sp}$ increases from 5\degr, over 20\degr~and 90\degr, to 180\degr. 
The spot co-rotates with the stellar surface ($v_{\rm sp}$=$v_{\rm rot}$), and $A_{\rm sp}$=0.5. 
The increase of $\Phi_{\rm sp}$ alters the FWHM evolution of the DAC with time. 
The base of the DAC strongly broadens because extra wind material injected by the spot is more distributed 
over the plane of the equator. 
The maximum density contrast $\rho/\rho_{0}$ inside the CIR therefore decreases when 
$\Phi_{\rm sp}$=90\degr~ increases to 180\degr~({\em lower panels}).
The CIR wind structure spreads out yielding broader DAC bases formed close to the stellar surface. 
Hatched areas are those with high Sobolev optical depth ({\em see text}).}
\label{fig Phisp effect}
\end{figure}

\begin{figure}  
\plotone{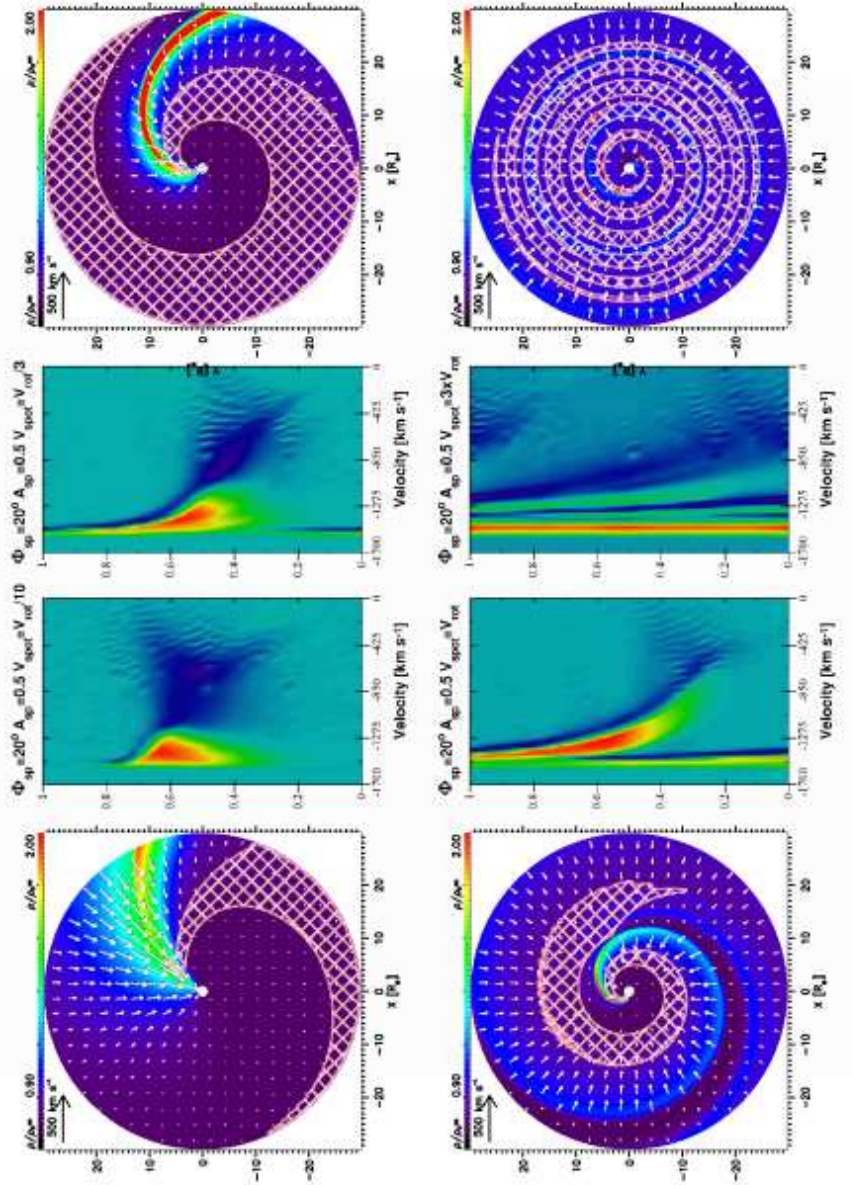}  
\caption{DACs are computed for spot rotation velocities $v_{\rm sp}$ increasing
from $v_{\rm rot}$/10, $v_{\rm rot}$/3 and $v_{\rm rot}$, to $v_{\rm rot}\times 3$. 
When the spot rotation lags behind the surface rotation the curvature of the CIR decreases, 
yielding DAC shapes that curve less over time ({\em upper panels}). 
Larger spot rotation velocities enhance the curvature of the CIR which further 
extends the spiral winding of the DAC line formation region around the star. 
For $v_{\rm sp}$=$v_{\rm rot}\times 3$ a single spiraling CIR yields several DACs 
crossing the line of sight at the same time ({\em lower right-hand panel}). 
Hatched areas are those with high Sobolev optical depth.}
\label{fig vsp effect}
\end{figure}

In Fig.~\ref{fig Phisp effect} we increase the spot opening angle from 5\degr, over 20\degr~and 90\degr, to 180\degr.
The spot co-rotates with the stellar surface, and $A_{\rm sp}$=0.5. 
We find that an increase of the spot angle considerably alters the FWHM evolution of the DAC. 
The DAC base strongly broadens when $\Phi_{\rm sp}$ increases from 20\degr~to 90\degr~because 
extra wind material injected by the spot becomes more spread out over the plane of the equator. 
The maximum of $\rho/\rho_{0}$ in the CIR occurs within $\sim$5 $R_{*}$ above the stellar surface. 
Inside this region the extra wind material is distributed over a larger geometric region above 
the bright spot (and surface), which also considerably broadens the density contrast in the tail of the CIR.  
When this region rotates in front of the stellar disk (around rotation phase 0.2 in the lower right-hand panel)
the DAC base becomes much broader because {\em the range of velocities projected in the observer's line of sight} 
(that contribute to the DAC opacity) strongly increases (see Sect. 2.2). Note that the 
wind flows almost radially through the CIR density structure near the surface (where particles decelerate as they approach and cross the CIR and
accelerate again once they have passed it) with rather small tangential components. 
It is however the range of projected radial velocities in front of the stellar disk 
that determines the total width of the DAC base. 
We find that the Sobolev approximation that only includes strictly radial velocity gradients 
yields line formation regions in the hydrodynamic models ({\em hatched regions in Fig.~\ref{fig Phisp effect}}) 
that do not correspond with the actual DAC formation regions when the tangential wind velocity 
components cannot be neglected.

\subsubsection{Spot Rotation Velocity $v_{\rm sp}$ }

The winding of the large-scale structures in the hydrodynamic models (i.e. the number 
of turns of the spiral arm in the equatorial plane around the central star) 
largely depends on the spot velocity (for fixed $v_{\infty}$). 
The spiral winding increases when the spot velocity 
increases with respect to the surface rotation velocity. 
Much smaller effects on the spiral winding are due to the spot angle; 
the spiral winding decreases somewhat when the spot angle increases. 

In Fig.~\ref{fig vsp effect} we compute the DAC for a spot rotation velocity $v_{\rm sp}$ that
increases from $v_{\rm rot}$/10, $v_{\rm rot}$/3 and $v_{\rm rot}$, to $v_{\rm rot}\times 3$,
with $A_{\rm sp}$=0.5 and $\Phi_{\rm sp}$=20\degr. When the spot rotation trails 
the surface rotation, the curvature of the CIR considerably decreases ({\em upper panels}).
For $v_{\rm sp}$=$v_{\rm rot}$/10  the density contrast in the CIR has the shape of a 
somewhat curved sector in the equatorial plane. 
When this sector rotates through the observer's line of sight (around rotation phases of 0.6-0.7), 
$|\frac{dv}{dr}|$ becomes small over a wind region that extends from the base of the wind 
(close to the surface) to the outmost wind regions beyond 30 $R_{*}$ in the hydrodynamic model 
({\em hatched area in the upper left-hand panel}). 
The DAC forms over the entire wind region along the line of sight which samples 
nearly all outflow velocities of the accelerating wind. 
The DAC has therefore an almost horizontal shape around rotation phases of 0.6-0.7 in the 
upper left-hand dynamic spectrum. 
An increase of the spot rotation velocity enlarges the curvature of the CIR, and further extends 
the spiral winding of the DAC line formation region around the star. 
The larger CIR curvature reduces the range of wind velocities the DAC samples along lines of sight 
where the radial wind velocity gradient is small and tangential wind components in front of the 
stellar disk can be neglected. 
The computed DAC therefore narrows towards larger $v_{\rm sp}$, and also appreciably alters 
its overall flux shape (e.g. the velocity position of the DAC minima with rotation phase). 
When the spot rotation leads the surface rotation (e.g. $v_{\rm sp}$=$v_{\rm rot}\times$3 in the 
lower right-hand panels), the curvature of the CIR becomes so large that the density spiral
warps several times around the central star. 
The large curvature of the CIR yields several DACs at any rotation phase that move simultaneously 
at different velocities through the dynamic spectrum, but which in fact all belong to the same 
large-scale spiraling wind structure produced by a single bright spot at the stellar surface.    

Besides introducing a parameter $v_{\rm sp}$ $\neq$ $v_{\rm rot}$, we also
considered the option of simply reducing the value of $v_{\rm rot}$
(which would of course be in contradiction with the observed
$v$~sin~$i$, but would provide the correct period of the DACs).
However, changes of $v_{\rm rot}$ would alter the smooth wind
due to the effect of the centrifugal acceleration.
As HD~64760 is rotating at 0.65 of the breakup velocity
\citep{Kaufer+al06}, a change of $v_{\rm rot} = 265\,\rm km\,s^{-1}$
to $100\,\rm km\,s^{-1}$ would increase $v_{\infty}$ by
$\sim$30\%, and would decrease the mass-loss rate by $\sim$30\%
\citep[see, e.g.,][]{Blomme91}. The changes in the underlying smooth wind
in turn would influence the development of the CIR. A model with
a lower $v_{\rm rot}$ is therefore not equivalent to one
with $v_{\rm sp}$ $<$ $v_{\rm rot}$.

This point is also relevant to the conclusion of \cite{Hamann01}
that (for spots locked onto the stellar surface) the DAC acceleration 
does not depend on the stellar rotation rate. The centrifugal acceleration
forces already influence the smooth wind structure, thereby creating a dependence on the rotational velocity.
One could of course artificially change the CAK parameters to compensate 
for the centrifugal acceleration in the smooth wind. But this would not 
compensate perfectly what happens inside the CIR, where the hydrodynamics 
is driven by a stronger radiation field. A (small) effect due to a different 
rotational velocity would therefore remain.

We summarize this Section by noting that the spot strength $A_{\rm sp}$ chiefly determines 
the density contrast inside the large-scale equatorial wind structures of hydrodynamic models.
Our 3D radiative transfer calculations reveal that $A_{\rm sp}$ 
mainly influences the velocity extension of the DACs in the absorption portion of the P Cygni wind profile. 
The spot opening angle $\Phi_{\rm sp}$ chiefly determines the width of the spiraling 
wind structures, which determines the FWHM evolution of the DAC over time (DAC morphology), 
and primarily the velocity range and shape of the DAC at its base. 
The rotation velocity of the spot $v_{\rm sp}$ with respect to the stellar surface rotation 
velocity mostly determines the curvature of the hydrodynamic structures, which mainly alters 
the velocity positions of the DAC flux minima over time (detailed DAC shape). 
Since the large-scale density- and velocity-structures in the hydrodynamic models are 
uniquely determined by the three spot parameters $A_{\rm sp}$, $\Phi_{\rm sp}$, and $v_{\rm sp}$
(assuming the smooth-wind properties are a priori known), they also uniquely determine the 
detailed DAC shape and morphology in our 3D radiative transfer calculations. 
It justifies to constrain the spot parameters from best fits to the detailed shape of 
observed DACs in the next Section.

\section{Best Model Fit}\label{dacfit}

In this Section we determine the spot parameters at the stellar surface of HD~64760 by matching 
the detailed DAC evolution in the Si~{\sc iv} lines. The best-fit spot parameters determine the density
contrast and flow velocities in the large-scale equatorial wind structures. The additional
mass-loss rate due to the CIRs
is therefore determined by the best-fit spot parameters, 
which we discuss in Sect.~\ref{discuss}.       

In Sect.~\ref{obs} we obtain a recurrence period of 10.3$\pm$0.5~d for the DAC observed in the Si~{\sc iv} lines of 
HD~64760. We assume that the star is observed edge-on (sin $i$=1). The surface rotation velocity of 
$v_{\rm rot}$=265 $\rm km\,s^{-1}$ 
yields a rotation period of 4.12~d (for $R_{*}$=22~$R_{\odot}$), 
or 2.5 times shorter than the observed DAC recurrence period. 
This is direct evidence that the spots cannot rotate with the surface.
We therefore use models having either one, two, four, or eight bright spots, 
with the spot parameter $v_{\rm sp}$ set equal to $v_{\rm rot}$ / 2.5, $v_{\rm rot}$ / 5, 
$v_{\rm rot}$ / 10, and $v_{\rm rot}$ / 20, respectively, in the equatorial structured wind models.
We compute the dynamic spectra of Si~{\sc iv} $\lambda$1395 in HD~64760 over a period of 15.5~d for direct 
comparison to the IUE observations. 

\subsection{One-spot model fits}\label{one spot fit}

\begin{figure}  
\vspace*{-0.7cm}
\plotone{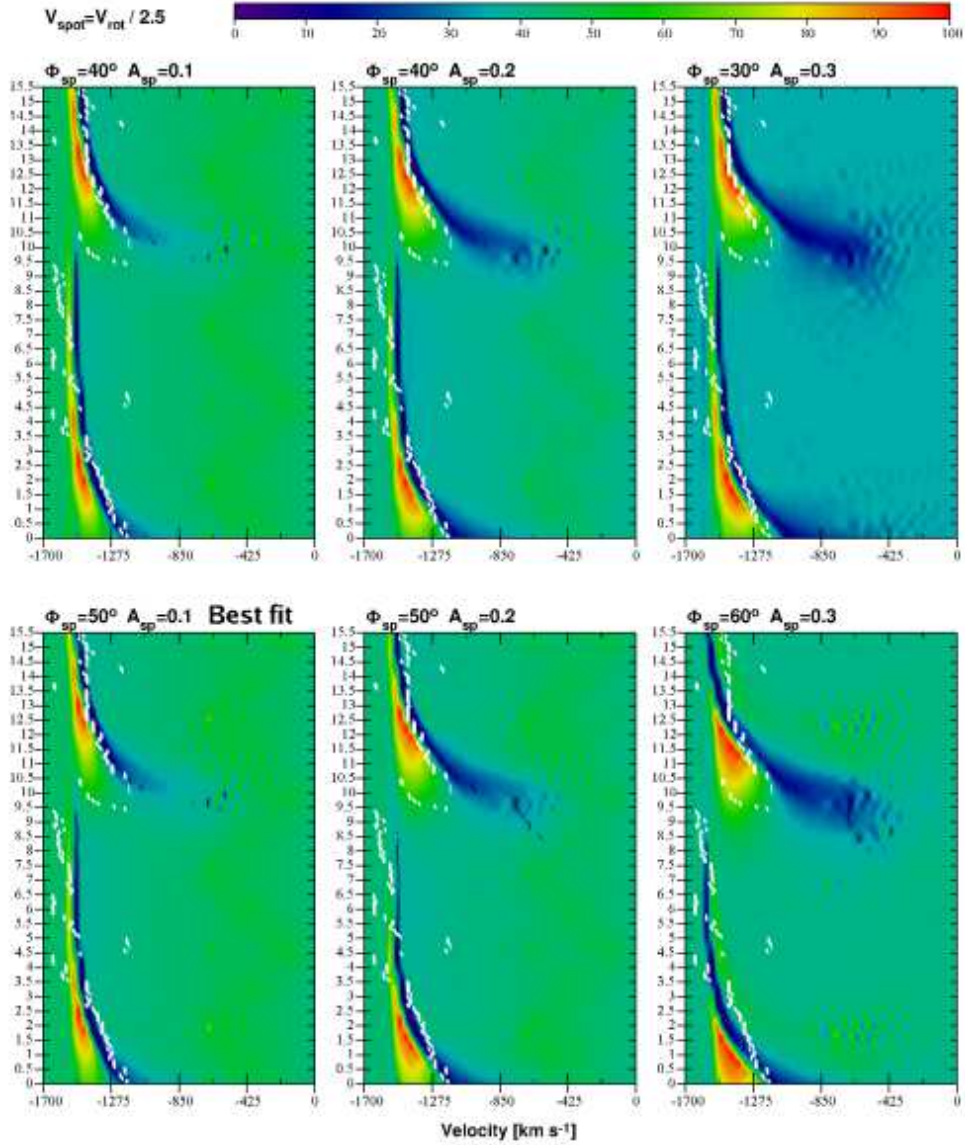}  
\caption{Part of an atlas of dynamic spectra 
(one-spot models)
of Si~{\sc iv} $\lambda$1395 in HD~64760 computed 
over a period of 15.5 d for various $A_{\rm sp}$ and $\Phi_{\rm sp}$ values that determine 
the detailed shape and morphology of the DAC. 
The spectra are shown between 0 and $-$1700 $\rm km\, s^{-1}$ and subtracted for the underlying 
smooth P Cygni wind profile. The flux minima in the DAC model are marked with black dots. The 
minima in the observed DAC are over-plotted with white dots. The hydrodynamic models 
are computed with a single bright spot that lags 2.5 times behind the surface rotation,
yielding an observed DAC recurrence time of 10.3~d. An increase of $A_{\rm sp}$ from 0.1 to 0.3 
({\em upper panels}) extends the DAC base toward unobserved velocities redward of $\sim$ $-$1000 
$\rm km\,s^{-1}$. $A_{\rm sp}$ therefore does not exceed 0.1. A least-squares minimization method 
applied to the observed and computed DAC flux minima ({\em black and white dots}) yields the best 
fit ({\em lower left-hand panel}) for one-spot models with $\Phi_{\rm sp}$ around 50\degr~({\em see text}).
}
\label{fig variants best fit}
\end{figure}

Figure~\ref{fig variants best fit} shows part of
an atlas of one-spot models. The Figure shows
six dynamic spectra computed with {\sc Wind3D} which are
closest to the best-fit solution.
The Figure contains certain combinations for
spot parameters $A_{\rm sp}$=0.1, 0.2, \& 0.3, and $\Phi_{\rm sp}$=30\degr, 40\degr, 50\degr, \& 60\degr.

We first discuss the goodness of fit by considering how well
the position of the model flux minima match the observations.
The computed spectra are shown between 0 and $-$1700 $\rm km\, s^{-1}$,
with the underlying smooth P Cygni wind profile subtracted. The computed flux minima in the DAC are
marked with black dots, while the flux minima of the observed DAC are over-plotted with white dots. 
We vary the spot parameters $A_{\rm sp}$ and $\Phi_{\rm sp}$ until a best match between the black and 
white dots is obtained. When we increase $A_{\rm sp}$ from 0.1, and 0.2, to 0.3 in the upper panels 
of Fig.~\ref{fig variants best fit}, the base of the DAC extends toward velocities redward of $\sim$ $-$1000 $\rm km\,s^{-1}$ 
that are
not observed in HD~64760 (Sect.~\ref{obs}). We therefore find that $A_{\rm sp}$ does not significantly exceed 0.1.
For $A_{\rm sp}=0.1\pm$0.05, the morphology of the computed DAC compares 
very well to the observed DAC, and $\Phi_{\rm sp}$ can be determined from a best fit to the detailed velocity 
position of the DAC flux minima over time. We apply a least-squares minimization method to the observed 
and computed DAC flux minima ({\em black and white dots}) yielding a best fit for models with 
$\Phi_{\rm sp}$ between 40\degr~and 50\degr. When $\Phi_{\rm sp}$=60\degr~({\em lower right-hand panel}) 
the computed DAC base crosses the observed DAC base too rapidly. If  $\Phi_{\rm sp}$ is lowered 
to 50\degr~(at most) the velocity positions of the flux minima at the DAC base shift bluewards almost linearly 
over time ({\em lower left-hand panel}) conform with the observed DAC evolution. For $A_{\rm sp}$ $<$ 0.1 
and $\Phi_{\rm sp}$ $<$ 40\degr~the DAC becomes invisible 
against the underlying smooth wind absorption.     
Our best-fit solution 
with a one-spot model
is therefore: $v_{\rm sp}$ = $v_{\rm rot}$ / 2.5;
$A_{\rm sp}=0.1 \pm 0.05$ and $\Phi_{\rm sp}$ = $50 \pm 10$\degr.

Figure~\ref{fig comparison best fit} compares the best fit dynamic spectrum  
({\em upper left-hand panel}) with the observed spectrum ({\em upper right-hand panel}). 
The computed dynamic spectrum correctly fits the observed flux difference spectrum in great detail. 
The velocity positions of the flux minima in the computed DAC ({\em marked with red dots}) 
differ by less than $\sim$50 $\rm km\,s^{-1}$ from the observed velocity positions ({\em white dots}) 
for $T$ between 0 and 3.5~d (lower DAC), and for $T$ between 10 and 15.5~d (upper DAC). 
Over these time intervals the computed flux evolution in the DAC base correctly fits the detailed 
shape and morphology changes of the observed spectra. 

We proceed with a detailed comparison of the observed DAC {\em shape} and the models.
In Fig.~\ref{fig detailed comparison best fit} we compare the computed ({\em left-hand panel}) 
and observed ({\em right-hand panel}) shape and morphology at the DAC base for 0~d $\leq$ $T$ $\leq$ 3.75~d. 
The dynamic spectra are plotted between $-$758 and $-$1666 $\rm km\,s^{-1}$, revealing the slanted 
triangular shape of the DAC base. The characteristic shape results from wind regions 
within a few $R_{*}$ above the stellar surface where the density increase of the CIR spreads out 
above the spot. The DAC line formation region rotates in front of the stellar disk and samples a 
decreasing range of (projected) radial wind velocities along the line of sight
causing a narrower DAC base over time. 
The FWHM of the computed DAC decreases from $\sim$100 $\rm km\,s^{-1}$ at $T$=0~d 
to $\sim$20 $\rm km\, s^{-1}$ around $T$=3.5~d, consistent with the observed DAC narrowing. 
The DAC width is small and stays nearly constant over the next 6.5~d, after which it fades away. 
The `tube-like' extension of the DAC base also corresponds to the observations (upper right-hand panel of 
Fig.~\ref{fig comparison best fit}). 
It results from the CIR above $\sim$20 $R_{*}$ in the DAC line formation region where 
the wind velocity is within 5\% of $v_{\infty}$=1500 $\rm km\,s^{-1}$. 
The DAC samples only a very small range of wind velocities around $v_{\infty}$, although 
its formation region (in front of the disk) extends geometrically from $\sim$20 to 30 $R_{*}$. 
The DAC approaches $v_{\infty}$, remains very narrow and visible, until the outer wind regions 
in the CIR rotate out of the observer's line of sight. 
The time scale over which the DAC remains visible is set by the curvature of the CIR around 
the star which mainly depends on $v_{\rm sp}$. 
Note that the computed DAC fades away after $T\simeq$ 10 d in the best fit model, as is observed in HD~64760. 
However, the velocity positions of the tube-like DAC feature observed between $T$= 7 
and 10~d somewhat exceed the DAC velocities of our best fit model by 100-150 $\rm km\,s^{-1}$. 
The smooth wind model has $v_{\infty}$=1500 $\rm km\,s^{-1}$, which we slightly vary to improve the fit. 
We find however that changing $v_{\infty}$ by $\pm$200 $\rm km\,s^{-1}$ leaves the spot parameters 
practically unchanged since the best fit criterion primarily relies on the DAC shape and 
morphology between $T$=0~d and 7~d.

\subsection{Two-spot model fits}\label{two spot fit}

\begin{figure}  
\vspace*{-0.9cm}
\plotone{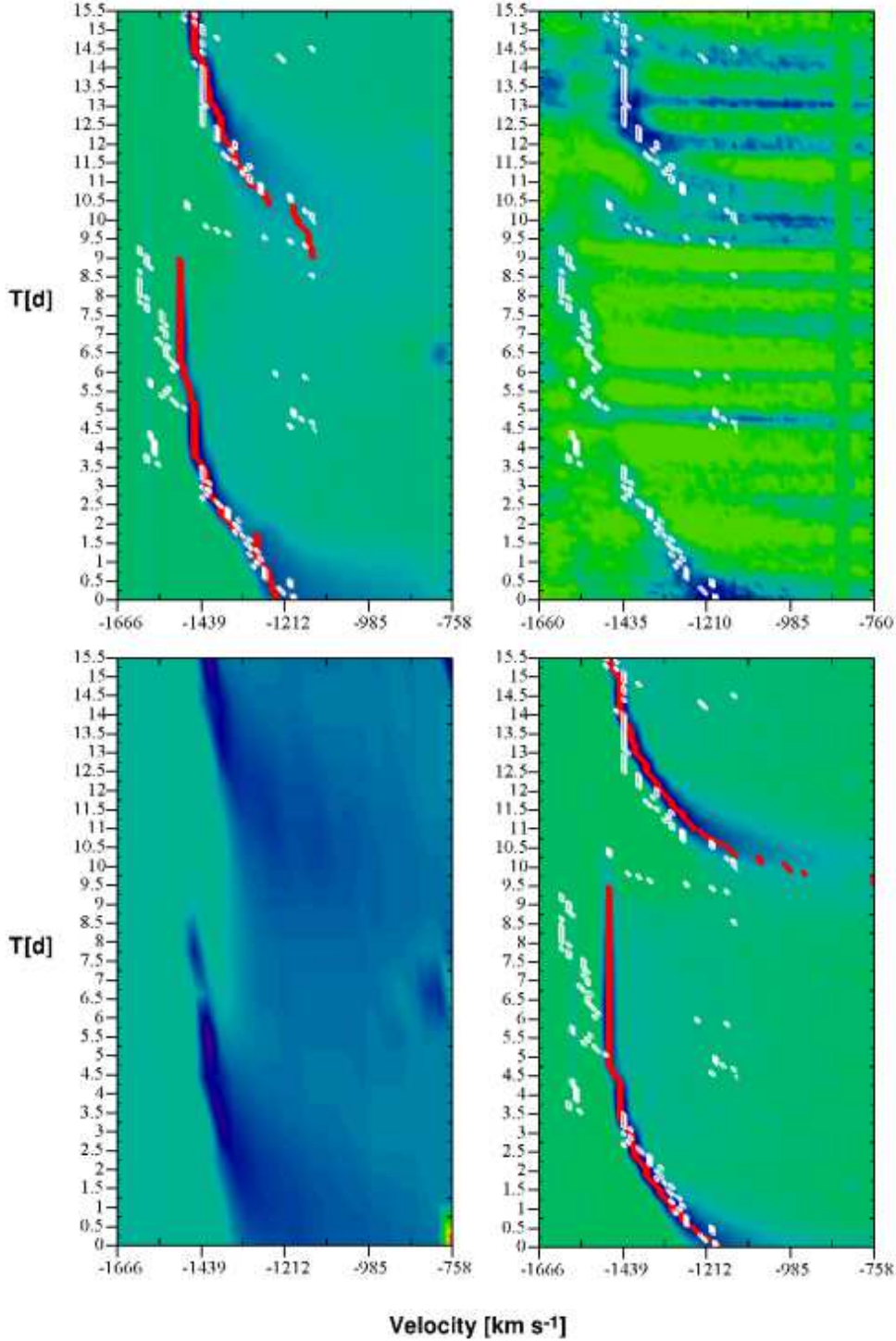}  
\caption{The best fit dynamic spectrum of Si~{\sc iv} $\lambda$1395 
for the two-spot model ({\em upper left-hand panel}) and the one-spot model ({\em lower right-hand panel}) 
compared to the observed spectrum ({\em upper right-hand panel}) of HD~64760. The shape and morphology 
of the computed DACs fit the properties of the observed DACs. The velocity positions of the 
DAC flux minima computed with the best fit two-spot model
differ by less than $\sim$20 $\rm km\,s^{-1}$ from the observed velocity positions ({\em white dots}) 
for 0~d $\leq$ $T$ $\leq$ 3.5~d, and 10~d $\leq$ $T$ $\leq$ 15.5~d.   
The FWHM of the computed DAC decreases from $\sim$100 $\rm km\,s^{-1}$ at $T$=0~d 
to $\sim$20 $\rm km\, s^{-1}$ around $T$=3.5~d, in agreement with the narrowing of the observed DAC. 
The lower DAC width remains almost constant over the following 6.5~d, after which it fades away. 
The `tube-like' extension of the DAC base is also observed in the upper right-hand panel, but somewhat 
exceeds the computed DAC velocities after $T$=7~d. A four-spot model in the lower left-hand panel 
yields DAC shapes and overlap not compatible with the observed shapes (see text).}
\label{fig comparison best fit}
\end{figure}

\begin{figure}  
\vspace*{-0.9cm}
\plotone{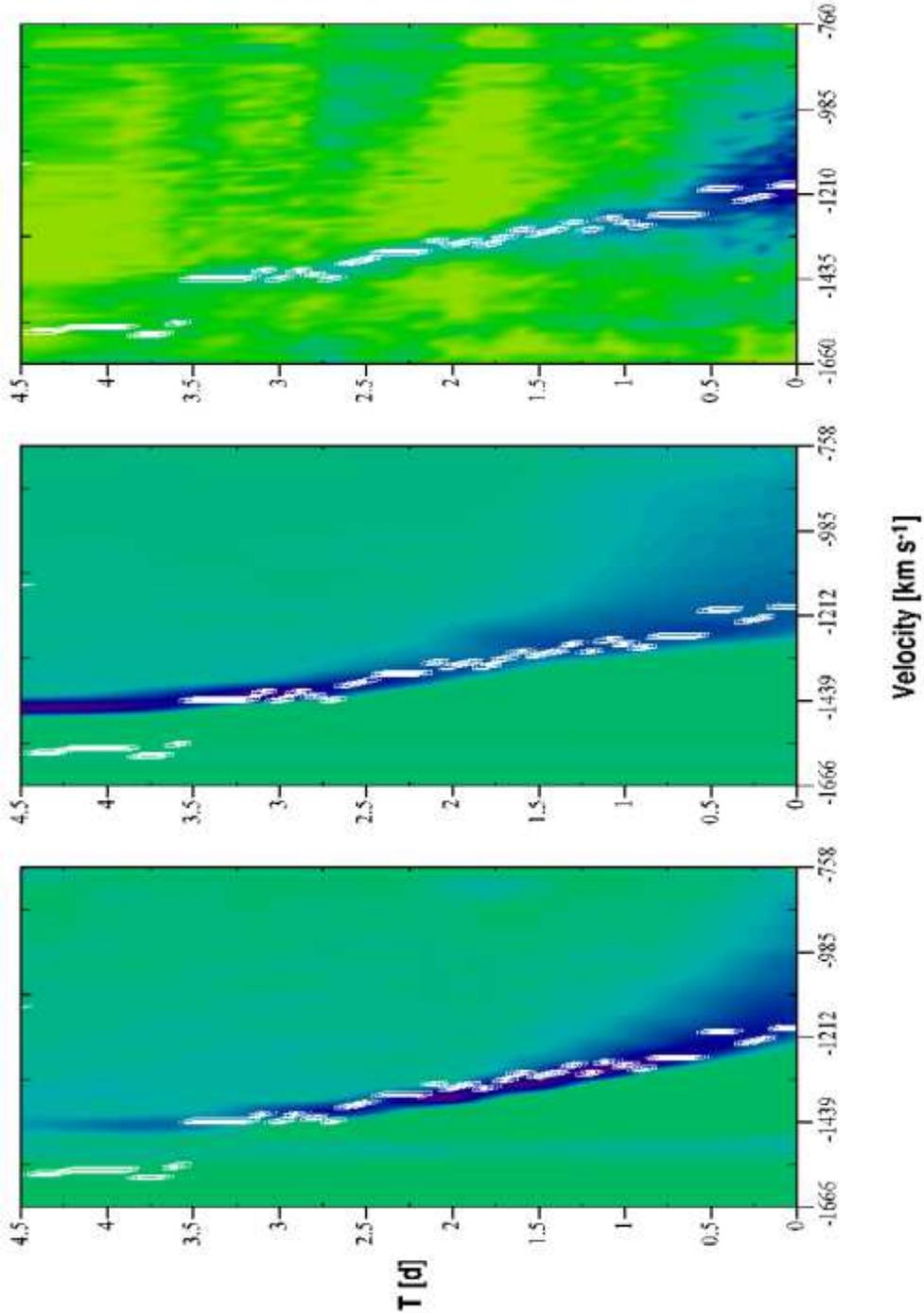}  
\caption{Detailed comparison of the shape and morphology of the DAC base 
for 0~d $\leq$ $T$ $\leq$ 4.5~d, computed with 
the best fit one-spot ({\em left-hand panel}) 
and the two-spot model ({\em middle panel}), with Si~{\sc iv} observations ({\em right-hand panel}) 
shown between $-$760 and $-$1600 $\rm km\,s^{-1}$. The slanted triangle of the DAC base emerges from wind regions
within a few $R_{*}$ above the stellar surface in Fig.~\ref{fig best fit}. The DAC line formation region 
rotates in front of the stellar disk and samples a decreasing range of wind velocities projected in the 
line of sight, yielding the narrowing of the DAC base over time. The decrease of the computed DAC 
width is strongly dependent of the three spot parameters since they uniquely determine the large-scale 
density- and velocity-structures the CIR produces in the smooth ambient wind. 
The one-spot dynamic spectrum is computed for an inclination angle $i$=85\degr~. 
The two-spot model with half the spot velocity of the one-spot model better fits 
the rather linear shape observed for the lower DAC 
({\em see text}).}
\label{fig detailed comparison best fit}
\end{figure}

The upper left-hand panel of Fig.~\ref{fig comparison best fit} shows the best fit for a 
two-spot model with $v_{\rm sp}$ = $v_{\rm rot}$ / 5. Models with two unequal spots 
better match the detailed DAC shape compared to our one-spot model fits. 
The spot parameters are varied separately until the best fit to the detailed shape 
and morphology of the observed upper and lower DAC ({\em upper right-hand panel}) is 
accomplished. We find the best fit to the lower DAC with a spot of 
$A_{\rm sp}=0.2 \pm 0.05$ and $\Phi_{\rm sp}$ = $20 \pm 5$\degr. 
The upper DAC
is best fit with a second spot inserted on the opposite side of the stellar equator
with $A_{\rm sp}=0.08 \pm 0.05$ and $\Phi_{\rm sp}$ = $30 \pm 5$\degr.
Since the spot velocity is halved compared to one-spot models, the curvature of
both CIR wind structures in Fig.~\ref{fig best fit} diminishes, yielding DAC shapes 
that are less curved. The left-hand and middle panels of 
Fig.~\ref{fig detailed comparison best fit} show the best fit to the detailed 
shape of the lower DAC with one- and two-spot models. The best two-spot fit 
to the lower DAC is an improvement and the computed flux minima correctly 
match the almost constant drift observed for its flux minima over time. 
The shape of the upper DAC flux minima is somewhat more curved (but less 
sharply defined due to the intersecting modulations), while its base is 
slightly more blue-shifted compared to the lower DAC.
The upper DAC is therefore best fit using a somewhat larger 
$\Phi_{\rm sp}$ = 30\degr. The larger spot opening angle used for the 
upper DAC broadens the computed width at its base and shifts the velocity positions of 
the flux minima very close to the observed values. The velocity 
extension of the upper DAC base somewhat decreases by reducing $A_{\rm sp}$ to 
0.08 compared to the lower DAC (see also Sect.~\ref{spot opening angle}).
One-spot models with $V_{\rm sp}$=$V_{\rm rot}$/2.5 produce DACs that accelerate 
considerably faster than a two-spot model with the same spot parameters, but having 
$V_{\rm sp}$=$V_{\rm rot}$/5. On the other hand, if the $V_{\rm sp}$ is 
also set identical for one- and two-spot models the differences in DAC 
shapes are very small and solely result from weak hydrodynamic interactions 
between the CIRs in the two-spot model (in the CIR tails at large wind velocity).                       

We also computed models with four and eight spots inserted at equal distances  
around the stellar equator. For models with four spots we set $v_{\rm sp}$=$v_{\rm rot}$ / 10,
with $A_{\rm sp}$- and $\Phi_{\rm sp}$-values for all four spots comparable to the two-spot best fit 
model parameters ({\em lower left-hand panel of Fig.~\ref{fig comparison best fit}}). 
We find however that hydrodynamic interactions between the four CIRs become 
stronger and substantially alter the resulting DAC structures. The high-velocity tail of the CIRs can 
partially overlap with the structures at the base of the wind from the trailing CIRs. The base of the 
upper DAC therefore time overlaps too long with the high-velocity tube-like extension 
observed in the lower DAC. The detailed morphologies of DACs computed with four- and eight-spot 
models are not sufficiently well separated over time to be compatible with the 
behavior of the DACs observed in HD~64760.                   

The best fit models are obtained from a least-squares fit to the 
minima of the lower DAC for a grid of one-spot models and a grid of two-spot models {\it separately}.
The least-squares fits to the lower DAC uses 36 spectra observed between 0 and 3.5 d 
because its DAC minima are poorly determined after 3.5 d (see Fig. 14). The least-squares 
fit procedure {\em within} both grids provides values that are significant 
indicators of the best fit spot parameters for the one-spot models or two-spot models. 
A least-squares fit comparison {\em between} one-spot and two-spot fits however 
is cumbersome strictly quantitatively because two-spot models always introduce 
two extra (spot) parameters in the fit compared to one-spot models. 
We compute a sigma (square root of the weighted sum of squares of velocity differences between the 
computed and observed DAC minima) of 22.2 $\rm km\,s^{-1}$ for the best fit two-spot model, 
and 24.7 $\rm km\,s^{-1}$ for the best fit one-spot model. These values are limited to 36 
spectra between 0 and 3.5 d, and can therefore be compared. 
The two-spot best fit model quantitatively better fits the observed 
minima than the one-spot best fit model. The difference of 2.5 $\rm km\,s^{-1}$ between 
these sigma values is however too small to conclude that the two-spot model 
significantly (in a statistical sense) better fits than the one-spot model 
when limited to the lower DAC. If the minima of the upper DAC are also included 
in the comparison, the difference of sigma values substantially increases, but 
the direct comparison of these values is more complicated because of the different number 
of free parameters.
          
We note however that none of the one-spot models yield the very linear shape 
we observe for the lower DAC. The one-spot models accelerate too fast, 
or the DACs are always too curved below the observed (rather straight) DAC shape. 
The two-spot models rotate twice slower and do provide this very linear 
DAC shape. The difference of DAC shapes in Fig. 14 is an
improvement of the best fit with one-spot models.
We also find that the two-spot best fit model yields a substantial 
broadening of the long-wavelength wing of the lower DAC base 
(between 0 and ~1 d), which better than the one-spot best fit model 
corresponds to the width changes observed in the lower DAC.        
It results from the smaller spot velocity in two-spot models.
The long-wavelength DAC base wing further broadens in the four-spot models
(Fig. 13) which rotate another factor of two slower. 
One-spot models leave no room to adjust for the observed shape of 
the upper DAC which is visibly (although perturbed) slightly 
different from the lower DAC. For example, in the lower right-hand 
panel of Fig. 13 the best fit one-spot model yields computed DAC 
minima that fall below the lower DAC and above the upper DAC. 
This fit problem is removed with the best fit two-spot model 
({\it upper left-hand panel}). 

Our best fit procedure does not include the detailed evolution of the DAC depth because the 
current hydrodynamics does not compute the extension of the models above and below the equatorial plane. 
The computed DAC depth is therefore determined by the geometric height $h$ of the CIR which
we set to 0.5 $R_{*}$ above and below the equatorial plane. An increase of $h$ of the structured 
wind model enhances the DAC depth in the P Cygni absorption profile because the amount of scattering 
due to the CIR in front of the stellar disk enlarges. 
We assume that the CIR height remains 
constant with distance from the star, having a fixed geometric height of $\pm$0.5 $R_{*}$
(the latter value is very close to the maximum height of a circular spot 
extending over $\pm$25\degr~above and below the equatorial plane 
we find for a best fit with one-spot models). 
The relative variation of the DAC depth over time is however dependent on 
the assumption of a fixed geometric height for the CIR model. The actual CIR height is of course expected 
to also increase with distance from the stellar surface (e.g. due to the 1/$r^{2}$ decrease of the
smooth wind density), causing the large-scale wind structures to grow 
in latitude over a `flaring angle' from the star. 
We however checked with 3D transfer calculations that $h$ and the flaring angle do not influence the detailed 
DAC shape or morphology we use for our best fits. In principle $h$ and the flaring angle can
be determined from fully 3D hydrodynamic model calculations. They would allow us to also invoke the detailed 
DAC depth changes in our best fit procedure. A calculation of more sophisticated 3D models 
is however beyond the scope of this paper. The DAC shape and morphology suffice to obtain the best 
spot parameters for large-scale equatorial wind structures.

While we assume in this paper that sin~$i = 1$ for HD~64760,
the inclination angle $i$ could be considered as
an additional free parameter in the 3D radiative transfer computations. 
The detailed model geometry ($h$ and flaring angle) of the structured wind influences 
the column densities that result from the CIRs along the line of sight
for an observer slightly above or below the equatorial plane.
Small flaring angles decrease
the relative DAC depth formed at the base of the wind (at small velocities in the dynamic spectra), 
while large inclination angles decrease the relative DAC depth formed in the outermost wind regions 
(at large velocities). Important influences of the observer inclination angle on the dynamic spectra are
shown in the left-hand panel of Fig.~\ref{fig detailed comparison best fit} with $i$=85\degr. 
Around $T$ = 3~d the DAC depth decreases rapidly because the CIR in wind regions above $\sim$10 $R_{*}$ 
is tilted below the stellar disk when viewed by the observer above the plane of the equator 
(compare to the lower right-hand panel of Fig.~\ref{fig comparison best fit} with $i$=90\degr). A larger 
decrease of $i$ decreases the DAC depth further. The DAC becomes invisible for 
$i$ $< $80\degr~when $h$ is constant and set equal to $\pm$ 0.5 $R_{*}$ in our 3D radiative transfer 
calculations. These calculations reveal that the observer line of sight towards HD~64760 does not exceed
$\pm$10\degr~from its equator-on direction, or that sin$i$ $>$ 0.984.

\section{Discussion}\label{discuss}

\subsection{Additional Mass Loss Rate}

At first sight, it may seem that we do not have enough information to evaluate
the mass-loss rate of the structured wind models because the hydrodynamics equations
are solved in the equatorial plane only. However, we do assume that the
spots are
circularly symmetric on the stellar surface, which we can use to determine 
the mass-loss rate. To first approximation, we can assume that very close to the 
stellar surface the wind density and radial flow velocity have the same 
circular symmetry as the spot. Hence, the mass-loss rate of the structured wind
($\dot{M}_{\rm struct}$) 
for a one-spot model
is:
\begin{equation}
\dot{M}_{\rm struct} = 2 \pi \int_0^{\pi}
   {\rm d} \theta' \sin \theta' \rho' (r,\theta') v' (r,\theta') r^2,
\label{equation Mdot struct}
\end{equation}
where $\theta'$ is the angle measured from the spot center. Transforming 
to
standard spherical coordinates $(r,\theta,\phi)$ gives
$\rho' (r,\theta') = \rho (r,\phi=\phi_0-\theta')$, where $\phi_0$
is the $\phi$ coordinate of the spot center. The assumption of circular 
symmetry in the wind is not exactly correct
however, not even close to the stellar surface. The leading edge of the density spiral 
slows down due to collision with the smooth wind, which results in a decrease of 
the mass-loss rate at the leading edge close to the surface. 
\citep[Note that the boundary conditions in the spot do not specify a 
       mass-loss rate, but rather use a constant-slope extrapolation
       for the radial velocity and a fixed base density -- see]
      []{Cranmer+Owocki96}.
To obtain a reliable estimate of the mass-loss
rate, we therefore neglect this effect by replacing the values of $\rho\, v$ smaller than
the smooth wind $\rho\, v$-values with the smooth wind values.

\begin{figure}
\plotone{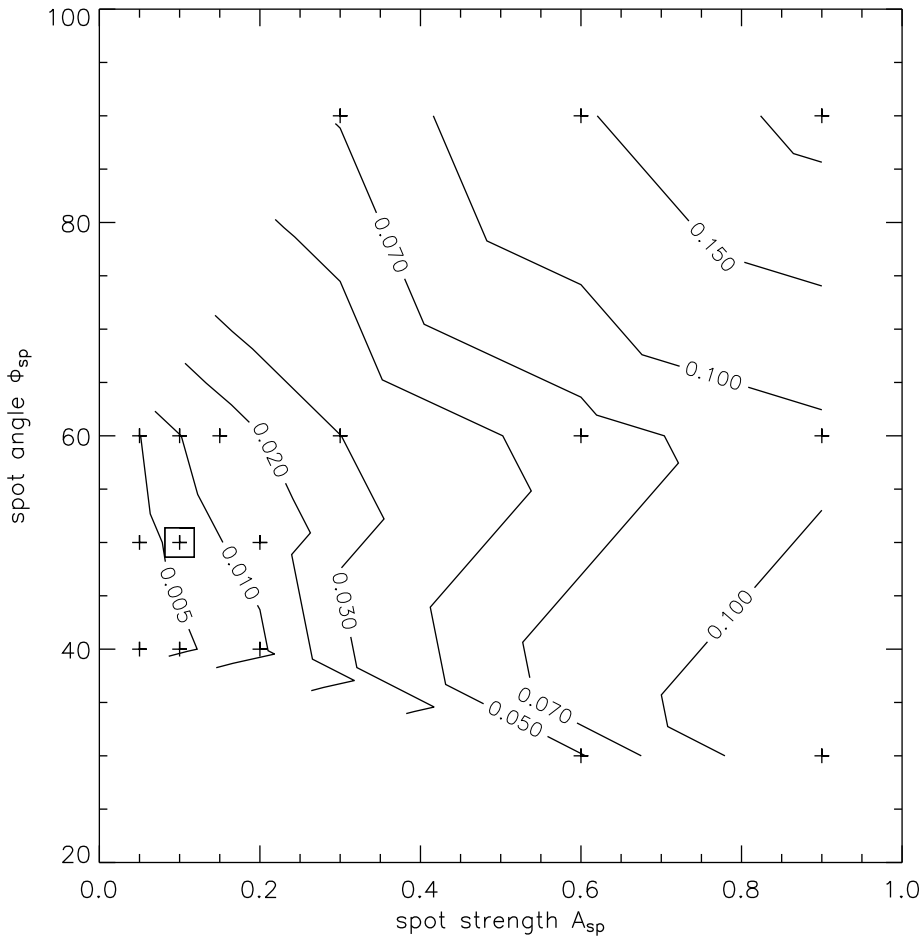}
\caption{Contour plot of the effect on the mass-loss rate due to the CIR (i.e.
$\dot{M}_{\rm struct}/\dot{M}_{\rm smooth}-1$), as a function of spot strength and 
size for a number of hydrodynamic models
with one spot.
The `+' symbols indicate computed models. 
The square marks the best-fit one-spot model for HD~64760. The mass-loss rate of 
the best-fit structured wind model with one spot is only 0.6 \% larger than the 
spherically symmetric smooth wind model.}
\label{fig Mdot effect}
\end{figure}

We use this method to compute the effect of the CIR on the mass-loss rate. 
In Fig.~\ref{fig Mdot effect} 
we plot $\dot{M}_{\rm struct}/\dot{M}_{\rm smooth}-1$
as a function of spot strength and spot angle for a number of 
one-spot
hydrodynamic models.
The Figure clearly shows that the extra mass-loss rate due to the CIR 
is always very small. Our best-fit model 
with one spot
in Sect.~\ref{one spot fit} yields an increase of 
only 0.6 \% for the mass-loss rate. 

The model mass-loss rates can be checked against a simplified version of 
Eq.~(\ref{equation Mdot struct}). 
In the model, the additional line force due to the spot is only
calculated for a position directly above the spot center, while for
other positions we relate it to that directly above the spot center
by using a Gaussian function of azimuthal angle 
(Sect.~\ref{section hydro models}).
If we simplify this by using 
a function that is unity above the spot
and vanishes outside it, we find
for a one-spot model
:
\begin{equation}
\frac{\dot{M}_{\rm struct}}{\dot{M}_{\rm smooth}} =
\frac{1}{2} \left( 1 - \cos \frac{\Phi_{\rm sp}}{2} \right)
\frac{\dot{M}_{\rm spot}}{\dot{M}_{\rm smooth}} +
\frac{1}{2} \left( 1 +  \cos \frac{\Phi_{\rm sp}}{2} \right),
\label{equation Mdot struct simple1}
\end{equation}
where $\dot{M}_{\rm spot}$ is the mass-loss rate over the spot.
We can further use the approximate relationship between mass-loss rate
and luminosity ($\dot{M} \propto L^{1/\alpha}$),
which casts Eq.~(\ref{equation Mdot struct simple1}) into:
\begin{equation}
\frac{\dot{M}_{\rm struct}}{\dot{M}_{\rm smooth}} =
\frac{1}{2} \left( 1 - \cos \frac{\Phi_{\rm sp}}{2} \right)
\left( 1 + A_{\rm sp} \right)^{1/\alpha} +
\frac{1}{2} \left( 1 +  \cos \frac{\Phi_{\rm sp}}{2} \right)
\label{equation Mdot struct simple2}
\end{equation}
We find very good agreement between the mass-loss rates derived from detailed 
hydrodynamic models with Eq.~(\ref{equation Mdot struct}) and the approximation of Eq.~(\ref{equation Mdot struct simple2}). 
This
demonstrates that our method of computing the mass-loss rates for structured 
wind models is appropriate.

Equation~(\ref{equation Mdot struct simple2}) can be extended to include
two symmetrical spots:
\begin{equation}
\frac{\dot{M}_{\rm struct}}{\dot{M}_{\rm smooth}} =
\left( 1 - \cos \frac{\Phi_{\rm sp}}{2} \right)
\left( 1 + A_{\rm sp} \right)^{1/\alpha} +
\cos \frac{\Phi_{\rm sp}}{2}
\end{equation}
Applying a similar formula to our best-fit model with two asymmetrical spots (Sect.~\ref{two spot fit}), we find a
mass-loss rate increase of only 0.5~\%.

\begin{figure}  
\plotone{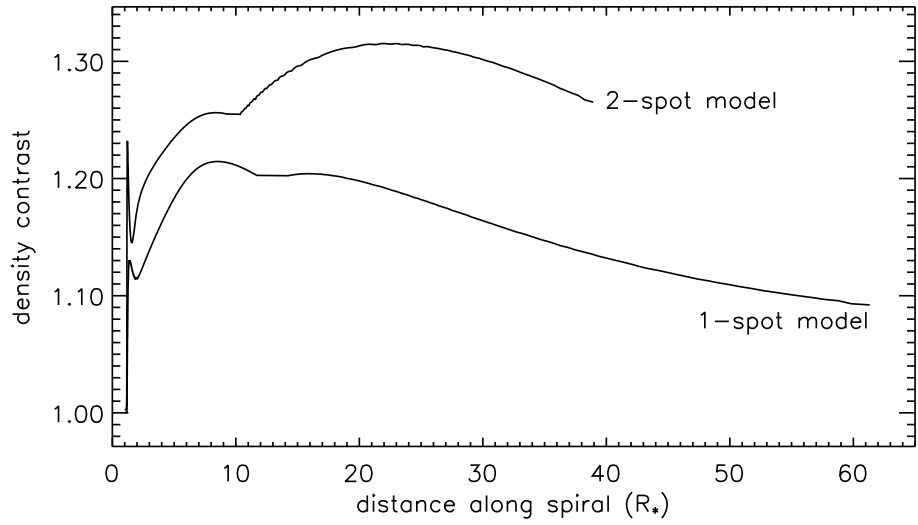}  
\caption{Density contrast $\rho$/$\rho_{0}$ along the CIR in the best-fit 
one-spot and two-spot hydrodynamic models
of HD~64760. The density in the equatorial CIR compared to the smooth 
wind density increases to a maximum of 
21\% (one-spot) or 32\% (two-spot).
}
\label{fig density contrast}
\end{figure}

Only
a relatively small enhancement of the mass-loss rate in the 
smooth wind is required to maintain the large-scale density- and velocity-structures in the equatorial 
plane of radiatively accelerated hot star winds. 
Figure~\ref{fig density contrast} shows the density contrast $\rho$/$\rho_{0}$
along the CIR in the best-fit one and two-spot hydrodynamic models of HD~64760.
The density enhancement 
in the CIR compared to the smooth wind density is only 32\% at most
(for the two-spot model). This type of
large-scale wind structure does not reveal the dynamic flow properties of a shock wave, but 
rather of a rotating density wave. The small density increase in the CIR signals that these 
large-scale extended wind structures can also easily be perturbed by other types of 
dynamic wind structures that may exist on smaller length scales in winds of massive hot stars. 
The best example would be the structure due to the instability of the radiative driving
mechanism \citep{Owocki+al88}.
The presence of DACs therefore provides a substantial constraint on the
amount of clumping present in the wind: too much clumping would
destroy the DACs \citep{Owocki98}.
The large-scale wind structures are a direct consequence of local 
irregularities in the radiative driving source at the base of the wind that involve limited 
variations of the surface intensity caused by dynamic surface structures that trail the stellar rotation.

\subsection{Comparison with Previous Work}

\citet{Kaufer+al06} studied the observed photospheric and H$\alpha$ line
profile variations of HD~64760. In the photospheric lines they detected
non-radial pulsations (NRPs) with three closely spaced periods
($P_1 = 4.810$~h, $P_2 = 4.672$~h, $P_3 = 4.967$~h, in order of 
decreasing amplitude).
To explain the variability in the H$\alpha$ line, they proposed a model
in which the beat period between $P_1$ and $P_2$ ($P_{\rm beat} = 6.8$~d)
is responsible for the CIRs in the wind. Both periods have quantum
numbers $l=-m$ and $\Delta m = 2$, so there are two diametrically opposed
spots on the equator. In general, the interference between modes
with arbitrary quantum numbers can result in considerably more
complicated interference patterns than is the case here. Because
$m_1 < m_2$, the beat pattern moves in the opposite direction of
rotation, with $-80.5~\rm km\,s^{-1}$. The whole pattern therefore
rotates in $4.12 \times 80.5/265 = 13.6$~d. Because there are two spots,
the effects on H$\alpha$ repeat with the beat period of 6.8~d.

A literal application of the \citet{Kaufer+al06} model to the IUE
data does not work, because the timescale over which their model
repeats (6.8~d) is not compatible with the observed 10.3~d timescale
in the UV DACs. This is the main reason why we did not use the
specific values from the \citet{Kaufer+al06} work, but only their
important idea of using spots that need not rotate with the rotational velocity.

Nevertheless, it should be noted that the beat period calculated
by \citet{Kaufer+al06} is quite sensitive to the accuracy with which
the NRP periods have been determined. A change in periods $P_1$ and $P_2$
of only 0.0235~h would result in a 10.3~d beat period, rather than
a 6.8~d one. From the power spectrum presented in their Fig.~5, 
it is clear that the NRP periods are not determined to such a precision.
We also note that their 6.8~d period is not compatible with the
suggested 2.4~d period in H$\alpha$. Interference between periods
$P_2$ and $P_3$ can give a 2.4~d beat period, provided that small shifts
to both periods are applied. It is therefore conceivable that the
UV DACs are due to a beat pattern between $P_1$ and $P_2$, while the
variability in H$\alpha$ and the UV modulations are due to
a beat pattern between $P_2$ and $P_3$. It of course remains to be explained why the $P_1$-$P_2$ beat pattern is only
seen in the UV DACs and the $P_2$-$P_3$ one only in Halpha.
The critical test is to
calculate both the Si~{\sc iv} and the H$\alpha$ line profile 
variations with our code. We defer this to a subsequent paper.

\section{Summary and Conclusions}\label{concl}

In this paper we show that a combination of advanced hydrodynamic and radiation transport calculations 
in three dimensions can correctly fit the detailed behavior of Discrete Absorption Components observed 
in variable UV wind lines of the hot supergiant HD~64760 (B0.5 Ib). The best fits are accomplished with 
sophisticated numerical simulations that invoke the physics of the transport of radiation in fast accelerating
stellar winds. The hydrodynamic models 
use spots of enhanced brightness on the stellar surface to initiate 
large-scale density- and velocity-structures rotating
in the plane of the equator.

We confirm and somewhat refine the conclusions of \citet{Cranmer+Owocki96}
regarding the effects of spot strength and opening angle on the resulting
CIRs. We also confirm their result that DACs are mainly formed in the
velocity plateau between the kink and the CIR. In some cases, however, 
the density enhancement is responsible for the DAC formation (e.g. at low velocities).
The conclusion of \cite{Hamann01} that (for spots locked onto the stellar
surface) the DAC acceleration does not depend at all on
the stellar rotation rate, cannot be extended. This is mainly due
to the effect of the centrifugal acceleration, which already plays
a role for the underlying smooth wind structure. Even when artificially
changing the CAK parameters to compensate for this, a (small) effect
would remain in the CIRs.

We find that the CIRs in fast-rotating HD~64760 are very extended 
density waves. The discrepancy between the 10.3~d recurrence time of the DACs 
and the estimated 4.12~d rotation period is direct evidence that spots cannot 
rotate with the stellar surface. For a two-spot model, they instead
lag 5 times (i.e. 2 $\times$ 10.3 / 4.12) behind the stellar surface rotation. The best fits to the observed DAC 
shape and morphology show that the Co-rotating Interaction Regions are caused by bright spots at the base 
of the wind that do not exceed the surface brightness by more than 20$\pm$5\%. The opening 
angles of the (circular) spots are around 20\degr~ and 30\degr~
diameter, 
hence together covering an appreciable fraction of the stellar surface area.      
The structured wind models reveal density enhancements that do not exceed the density of the 
surrounding smooth wind by more than $\sim$~30~\%. The wind flow radially crosses the CIRs in the plane of the 
equator where it decelerates by less than $\sim$~140 $\rm km\,s^{-1}$. 
The base of the DACs form in wind regions within a few $R_{*}$ above the surface where extra material 
spreads out above the spots. The width of the DACs decrease over time because the CIR 
structures rotate across the stellar disk, causing regions of larger optical depth 
farther out in the accelerating wind where the radial velocity dispersion in front 
of the stellar disk decreases. 
The properties (shape and dynamics) of the DAC line formation regions are uniquely determined 
by the spot properties at the base of the wind. It enables us to obtain a unique best fit to 
the shape and morphology of the DACs. We find 
a better fit to the shape of the upper and lower DACs with a
two-spot model than with a one-spot model.
Our detailed transfer calculations correctly match the (slanted) triangular shape of the DAC base  
that further extends into a (tube-like) bluewards-drifting narrow absorption feature observed 
in HD~64760 (and also in $\xi$~Per). 
The spots at the surface increase the total mass-loss rate of the smooth symmetric wind model for HD~64760
by only $\sim$0.5\%.

We conclude that the large-scale structured wind of HD~64760 is caused by perturbations of the 
surface intensity that trail the fast surface rotation. This points to dynamic structures at 
the base of the wind that do not co-rotate 
with the stellar surface
(unlike, e.g., surface magnetic fields).
Such structures could possibly result from an interference of 
non-radial pulsations traveling the stellar circumference with periods set by the properties of permitted 
stellar pulsation modes. The relatively small perturbations that are required to create these DACs
probably explains why they are so ubiquitous. 
Their presence also puts a substantial constraint on the amount of
clumping that can be present in the wind.
The large-scale coherent CIR structures may become perturbed by wind 
clumping on much smaller length scales. They will however built up again 
very rapidly as well, provided that the perturbation 
time-scales are sufficiently short for the CIR structures to completely 
develop and to cause the UV DACs.
In future work we will investigate the effect of CIRs on
other spectral lines and other stars, trying to best fit the observational data.
For stars not observed close to equator-on, we will develop 3D hydrodynamic models.

\acknowledgments
This work has been supported by the Belgian 
Federal Science Policy - Terugkeermandaten.
We thank Asif ud-Doula for making his version of the {\sc Zeus3D} code available
to us. We thank the referee for several comments that improved the content and clarity of the paper.

\appendix
\centerline{\bf APPENDIX}

\section{3D Transfer Formalism and Discretization}
We denote by the vector $\bf{n}$ the direction in which the light rays 
travel at position coordinates $\bf{p}$.
$I$=$I_{\nu}$($\bf{p}$,$\bf{n}$) denotes the specific intensity at 
frequency $\nu$ of radiation traveling in direction $\bf{n}$ through point $\bf{p}$ (or the 
direction of photon extinction at $\bf{p}$). We solve the transfer problem in the observer frame 
because the co-moving frame formulation is not applicable to structured winds that can expand 
asymmetrically and that have non-monotonic wind velocity structures. 
When we denote the opacity and emissivity of the material $\chi$=$\chi_{\nu}$($\bf{p}$,$\bf{n}$)
and $\eta$=$\eta_{\nu}$($\bf{p}$,$\bf{n}$), respectively, the time-independent monochromatic 
transfer equation at $\bf{p}$ for a ray traveling in direction $\bf{n}$ is $\bf{n}$ $.$ $\nabla$ $I$ = $\eta$ $-$ $\chi$ $I$.
We use the two-level atom approximation with complete frequency redistribution so that 
$S^{l}$ = (1-$\epsilon$) $\bar{J}$ + $\epsilon$ $B$, where $S^{l}$ = $S^{l}$ ($\bf{p}$) is the 
line source function, $B$ = $B$ ($\bf{p}$,$T$) the Planck function at $\bf{p}$, 
$\epsilon$ the 
`thermalization parameter' 
in the non-LTE transfer problem, and $\bar{J}$ the frequency-weighted mean 
intensity at $\bf{p}$. 
In this paper we solve the pure scattering problem for resonance lines formed in 
the very extended winds of massive stars that are considered to be isothermal in the 
lines formation region. In these conditions the 
spontaneous de-excitation rate greatly exceeds the collisional transition rate.
We therefore assume $\epsilon$ = 0, and currently neglect LTE contributions 
in the transfer problem.
It permits to solve the pure scattering transfer equation with the line source function at $\bf{p}$
\begin{equation}
 {\it S}^{l}({\bf p}) = {\bar J}({\bf p}) = \frac{1}{4 \pi} \int_{4\pi} \int_{0}^{\infty} 
\phi_{\nu}(\nu,{\bf n}) \, {\it I}_{\nu}({\bf p},{\bf n}) \, {\it d}\nu \, {\it d}\Omega \,,           
\end{equation}
which integrates the specific intensity over all spatial angles $\Omega$,
weighted by the line profile function $\phi_{\nu}$. $\theta$ and $\psi$ denote the two spherical 
coordinate angles with $d\Omega$=sin($\theta$)\,$d\theta$\,$d\psi$. 
Note that we use $\psi$ instead of the more common $\phi$, to avoid
confusion with the profile function. When we denote $\nu_{0}$ the 
rest frequency of the line and $\bf{v}$ the wind velocity vector at $\bf{p}$,
the Gaussian line profile function $\phi$ = $\phi_{\nu}(\nu,\bf{n})$ is  
\begin{equation}
\phi_{\nu}(\nu,\bf{n}) = \frac{1}{\Delta \nu_{D}\sqrt{\pi}} \, {\rm exp} \left( -\left(\frac{\nu-{\nu_{0}}{(1-\frac{v.n}{{\it c}})}}{\Delta \nu_{D}}\right)^{2} \right) \\.
\end{equation} 
$\Delta \nu_{\rm D}$ denotes the line width $\Delta \nu_{\rm D}$=$\nu_{0}$ $({v_{\rm th}^{2}+ v_{\zeta}^{2}})^{1/2}$/c, 
with $v_{\rm th}$ the mean thermal particle velocity and $v_{\zeta}$ the turbulence velocity. 
We assume a Gaussian profile function for lines that are not intensity saturated.
We do not consider line formation on the root part of the curve of growth, which would otherwise 
require parameterized damping profile functions. Also, we currently do not include a partial redistribution or 
correlation function in the atom's frame because the 3D transfer problem should remain tractable by 
assuming complete redistribution over all line frequencies.
Note that the integration of Eq.~(A1) yields a frequency independent line source function in $\bf{p}$, 
which renders the problem of lambda iteration on $S^{l}$ independent of detailed frequency redistribution 
in the line profile. 

The 3D scattering transfer problem consists of calculating $S^{l}$ by summing in Eq.~(A1) over contributions from 
light rays that travel through $\bf{p}$ from all directions, while accounting for Doppler 
effects in Eq.~(A2) caused by the local wind velocity $\bf{v}$, 
and lambda iterating till $S^{l}$ becomes consistent 
with the radiative transfer equation. With $\eta$=$\eta^{\rm c}+\eta^{l}$ and $\chi$=$\chi^{\rm c}+\chi^{l}$   
the isotropic contributions from the continuum radiation, and anisotropic contributions from the line radiation 
to $\eta$ and $\chi$, the transport equation for stationary advection of the radiation field with respect to $\bf{n}$ is
\begin{equation}
\bf{n}\, .\, \nabla {\it I} = \chi^{l} \, \, {\it S}^{l} + \chi^{c} \, {\it S}^{c} - (\chi^{l}\, + \chi^{c})\, {\it I} \,,
\end{equation} 
where $S^{\rm c}$=$S^{\rm c}$($\bf{p}$) denotes the continuum source function and $S^{\rm c}$=$\eta^{c}$/$\chi^{\rm c}$. 

The time-independent multi-dimensional radiative transfer problem 
involves the development of a sophisticated numerical scheme that has to be 
tailored to solve a Boltzmann equation for photons. A wide variety of such 
algorithms is discussed in the literature which, dependent on the considered astrophysical 
conditions, are suitable to solve this equation from its diffusion limit 
(optically thick) to the pure transport (optically thin) limit. In
astrophysical line and continuum transport, where scattering cannot 
be neglected, one applies a finite element discretization 
technique to solve 
an integro-differential equation (or a set of them).
Generally, the complexity of this transfer problem results from the 
fact that due to scattering the radiation field is not only determined 
by local gas conditions but has to be consistent with the (atomic) 
state of the matter through which it can propagate over large geometric 
distances. While for 1D problems numerical methods are well established 
\citep[e.g.,][]{Hubeny97}, adequate solutions for 3D problems are still in their infancy.            
A concise overview of 3D non-LTE radiative transfer problems in the past 
decades has recently been presented in \citet{Hauschildt06}.     
We opted to implement Adam's  \citep{Adam90} spatially implicit first order method because 
it is one of the few 3D schemes that has successfully been applied to spectral lines 
that form in astrophysical conditions such as Be-star disk winds \citep[e.g.,][]{Hummel94}.

Adam's method consists of integrating Eq.~(A3) for a finite volume in a Cartesian system 
of coordinates. The computation of $I$ and $S^{l}$ involves solving a system of integro-differential 
equations (Eqns. (A1) \& (A3)) for which we consider an equidistant 3D rectangular grid with grid-points
$\bf{p}$=$\bf{p}_{\it ijk}$=($x_{i}$,$y_{j}$,$z_{k}$) where $i$=1..$N_{x}$, $j$=1..$N_{y}$, $k$=1..$N_{z}$, and
with velocity vectors $\bf{v}$= $\bf{v}_{\it ijk}$=($v_{x_{i}}$,$v_{y_{j}}$,$v_{z_{k}}$). Along a chosen direction $\bf{n}$, 
Eqn. (A3) can be discretized in the observer frame, and cast into
\begin{eqnarray}
I_{ijk}=\left( \chi^{l}_{ijk} \, S^{l}_{ijk} + \chi^{c}_{ijk} \, S^{c}_{ijk} +
n_{x_{i}} \, \frac{I_{i-\alpha,j,k}}{x_{i}-x_{i-\alpha}} + n_{y_{j}} \, \frac{I_{i,j-\beta,k}}{y_{j}-y_{j-\beta}} +
n_{z_{k}} \, \frac{I_{i,j,k-\gamma}}{z_{k}-z_{k-\gamma}} \right) /  \nonumber \\
\left( \chi^{l}_{ijk} + \chi^{c}_{ijk} + \frac{n_{x_{i}}}{x_{i}-x_{i-\alpha}} + \frac{n_{y_{j}}}{y_{j}-y_{j-\beta}} + \frac{n_{z_{k}}}{z_{i}-z_{i-\gamma}}       \right) \,.
\end{eqnarray} 
The monochromatic specific intensity $I_{ijk}$=$I_{\nu}(\bf{p}_{\it ijk},\bf{n}$) in $\bf{p}$ is computed with Eq.~(A4) by incrementing $\alpha$, $\beta$, and $\gamma$ by $+1$ or $-1$ depending on the direction $\bf{n}$ of the rays in the grid. 
A general application of the Adam's method treats $\chi^{\rm c}$ as a parameter, 
while $S^{c}$ it set equal to $B$ throughout. Since $B$ is constant in an isothermal 
wind over one line width, we can set $\eta^{\rm c}$ equal to a constant value. 
The detailed treatment of continuum processes is not included at this point.
For our application of Adam's method in this paper, we set $\chi^{c}$=0 and $S^{c}$=0 
throughout the wind, and only use non-zero values for the continuum flux emitted at the 
stellar surface.
We set the boundary conditions $\chi^{l}$=0 and $\chi^{c}$=0 for grid-points outside the grid. 
To be consistent with the 
nearly 
spherical geometry of a stellar wind, 
we also set these boundary conditions on the sphere that just fits inside our simulation volume.
The corners of the grid ($|\bf{p}_{\it ijk}|$ $\geq$ $R_{\rm max}$)
therefore also have $\chi^{l}$=0 and $\chi^{c}$=0.
At the boundary of the grid
we set $I_{ijk}$=0 and solve Eq.~(A4) by stepping through the indices {\it i,j,k}.
This is done for a set of observer frequencies $\nu_{a}$ and a number of directions
specified by angles ($\psi_{b}$,$\theta_{c}$). We then
compute $S^{l}_{ijk}$ or $\bar{J}_{ijk}$=$\bar{J}_{ijk}$($\bf{p}_{\it ijk}$) with a quadrature sum over $I_{ijk}$;
\begin{equation}
\bar{J}_{ijk} = \frac{1}{4\pi} \sum_{a=1}^{N_{\nu}}A_{a}\,\sum_{b=1}^{N_{\psi}}B_{b}\, \sum_{c=1}^{N_{\theta}}C_{c}\,\,{\rm sin}(\theta_{c})\,\,I_{ijk}(\nu_{a},\psi_{b},\theta_{c})\,\,\phi(\nu_{a},{\bf{v}}_{\it ijk},\psi_{b},\theta_{c})\,.
\end{equation}         
The summation in Eq.~(A5) can be limited to an interval (1..$N_{\nu}$) around the rest frequency $\nu_{0}$ that covers 
the entire wind line (i.e. with a P Cygni profile). The summation over angular coordinates 
($\theta_{c}$,$\psi_{b}$) is computed with a number of integration points that decreases toward higher latitudes $\theta_{c}$ on the sphere
by using $N_{\psi}$=$N_{\psi}$($\theta_{c}=\pi/2$)\,{\rm sin}($\theta_{c}$).
The calculation of $\bar{J}_{ijk}$ therefore sums 
angles (or directions $\bf{n}$) that are isotropically distributed around ${\bf{p}}_{ijk}$. It ensures that 
the 3D transfer is correctly solved for random opacity and velocity distributions (i.e. no 
preferential directions are considered for transfer in for example an equatorial disk). Note in Eq.~(A1) that 
both $I$ and $\phi$ are direction dependent which makes the accuracy of Eq.~(A5) very sensitive 
to the total number of directions $\bf{n}$ in the summation. The number of $\bf{n}$ (or $N_{\psi}$,$N_{\theta}$)
in Eq.~(A5) scales with the size of the grid ($N_{x}$,$N_{y}$,$N_{z}$), because grid-points at increasing distances from 
${\bf{p}}_{ijk}$ require finer sampling of $\bf{n}$ to ensure the same accuracy of ${\bar{J}}_{ijk}$.     
The quadrature coefficients $A_{a}$, $B_{b}$, and $C_{c}$ in Eq.~(A5) are computed with the fast (and simple) 
three-point Simpson formula for uneven grid-sizes, while Simpson's 3/8 rule is used in case of even grid-sizes.        

Equations (A4) and (A5) are solved iteratively whereby $S^{l}_{ijk}$ converges to 
a solution that changes by less than 1\% over following cycles (exact lambda iteration). This technique 
is used to evaluate $S^{l}_{ijk}$ over successive iterations (iteration between radiation transfer and statistical 
equilibrium), followed by one final transfer iteration that computes the emergent line fluxes for a given 
set of directions $\bf{n}$. 
For the implementation of {\sc Wind3D} the transfer computations are separated into an `iteration phase' (discussed in Appendix~B)
and an `interpolation phase' (Appendix~C) because of memory limitation requirements and parallelization optimizations.  

\section{3D Lambda Iteration}

During the iteration phase {\sc Wind3D} computes the lambda iteration on $S^{l}_{ijk}$ in every grid-point starting 
from initial values that are well chosen as to limit the number of lambda iterations. The {\em starting} values
are computed with the Sobolev approximation in Eq.~(A4). The structured wind opacity $\chi^{l}$ in strong UV wind 
lines is computed from the wind density contrast since  
$\overline{\chi^{l}}$ $\simeq$ $\overline{\chi^{l}_{\rm smooth}}$ $\rho$/$\rho_{\rm smooth}$, where $\overline{\chi^{l}}$ 
is the frequency integrated line opacity
(i.e. $\chi^{l}$= $\overline{\chi^{l}}$ $\phi$)
The smooth wind opacity $\overline{\chi^{l}_{\rm smooth}}$ in the lines is given by the
\citet{Groenewegen89} parameterization:
\begin{eqnarray}
\overline{\chi^{l,ijk}_{\rm smooth}} & = & \frac{T_{\rm tot}}{Y_{ijk}} \left(\frac{v_{ijk}}{v_\infty}\right)^{\alpha_1} 
   \left[1 - \left(\frac{v_{ijk}}{v_\infty}\right)^{\frac{1}{\beta}} \right]^{\alpha_2}
   \frac{{\rm d}v_{ijk}}{{\rm d}r} \frac{1}{v_{\rm th}} \\
Y_{ijk} & = & \int_{v_0}^{v_\infty} \left(\frac{v_{ijk}}{v_\infty}\right)^{\alpha_1}
   \left[1 - \left(\frac{v_{ijk}}{v_\infty}\right)^{\frac{1}{\beta}} \right]^{\alpha_2}
   {\rm d}\left(\frac{v_{ijk}}{v_\infty}\right),
\end{eqnarray}
where $\beta$ is the parameter of a $\beta$ velocity law for the surrounding smooth wind. 
$v_{ijk}$ is the radial wind velocity in $\bf{p}$, and $v_{\rm th}$ the thermal broadening 
for the scattering ion. Note that this neglects the 
large turbulent broadening found by 
\citet{Hamann81} 
and \citet{Groenewegen89}.
$T_{\rm tot}$, $\alpha_1$ and $\alpha_2$ are the opacity parameters, and the other symbols have their
usual meaning. 
These opacity parameters determine whether or not the absorption portion of the 
computed profile intensity saturates. 
The values of the parameters are determined from 
an approximate fit of  
the computed average flux profile to the observed line profile.
The detailed shape of the computed underlying
line profile is not further optimized as we do not model the precise DAC 
depth in the profile, but rather its shape and FWHM evolution.    
The value of $\beta$ in Eqns. (B1) 
and (B2) is however determined from a best fit using a $\beta$-velocity law to 
the smooth wind structure of the hydrodynamic model. For our best fit hydro 
model of HD~64760 we find $\beta=0.71$.  

It is of note that Adam's scheme uses the widespread short characteristics 
method \citep[e.g.,][]{Koesterke+al02}. The method is rather straightforward to implement 
for solving the non-relativistic, two-level atom, 3D transfer problem 
with complete frequency redistribution, which includes the solid angle and 
frequency grid parallelization. We find that this is particularly the case 
for the pure scattering problem in optically thin (wind) line formation 
conditions we consider in this paper. Adam's method of lambda iterating the 
line source function to consistency with the transfer equation 
in a two-level atom
proves 
to be very efficient and avoids the usual problems with non-convergence. 
For the pure scattering problem ($\epsilon$=0) {\sc Wind3D} always converges 
$S^{l}$ within a dozen iterations at most. \citet{Adam90} already pointed 
out that the method of Cartesian upwind discretization is unconditionally 
stable because only positive terms occur in Eq.~(A4) (see also the study by \citet{Hummel92}
of the transfer scheme). Our combination of a starting Sobolev source 
function with Adam's discretization method provides an adequate lambda iteration 
scheme that is particularly suited for 3D scattering dominated transfer in 
asymmetric wind opacity- and velocity-structures. 

The short characteristics method is
however known to suffer from beam widening because the interpolation 
introduces angular diffusion in the numerical solution.  
The long characteristics method removes this defocusing problem because much 
higher accuracies can be achieved with it. The latter method is however
computationally very expensive if one wants to sample high angular resolution, so as 
to accurately sample space at large distances from the source in an extended structured wind.      
It is also time-consuming because the long characteristics usually cover the same
part of the transfer domain many times, which introduces strong redundancy. 
Long characteristics methods have only recently been examined within the framework 
of 3D radiative line transfer, but still await applications to real astrophysical problems 
\citep{Baron07}.

In our current implementation, Adam's method also requires large memory storage capacity 
during the iterations because the computation of $I_{ijk}$ in Eq.~(A4) requires to store two matrices for 
$I_{i-\alpha,j,k}$ of at least $N_{y}\times N_{z}$ elements for lightrays traveling in the 
forward and backward (incrementing $\alpha$ with $+$1 and $-$1) direction along the $x$-axis of the grid.  
This number has to be multiplied by the number of directions that is considered for the isotropic 
integration of $\bar{J}_{ijk}$ in Eq.~(A5) and the frequency resolution of the line profile, yielding
2$\,N_{y}\,N_{z}\,N_{\psi}\,N_{\theta}\,N_{\nu}$ matrix elements per iteration cycle. The requirement 
that the integration accuracy of $\bar{J}_{ijk}$ is maintained toward larger grid-sizes demands
that the number of directions $N_{\psi}\times N_{\theta}$ scales linearly with $N_{y}\times N_{z}$. 
For $N_{\nu} \simeq N_{y}$ we therefore find that the required memory capacity scales with $N^{5}$
for fast computation of $\bar{J}_{ijk}$. For example, a doubling of the size of a cubic grid 
requires an increase of total memory with at least a factor of 32. In principle the large memory 
requirements can be reduced by solving Eq.~(A4) monochromatically in one direction (per $\nu$ and per $\bf{n}$) 
whereby Eq.~(A5) is then computed in incrementing steps. The bottleneck of the 3D lambda iteration
is however in Eq.~(A5), which can adequately be parallelized by distributing the frequencies 
(and directions) over different threads (multi-threading). {\sc Wind3D} currently implements a 
parallelization strategy that 
provides fastest possible integration of $\bar{J}$. The parallelization technique of $\bar{J}$ yields 
excellent load balancing results with the least occupied CPU (using one thread per CPU) always performing 
within a few percent of the most occupied one.     

We also implemented 3D lambda iteration in {\sc Wind3D} that invokes an accelerator scheme to improve  
convergency times for the source function calculation. We use a linear average of source functions computed 
over preceding iterations cycles to predict the source function for the next iteration. In case of slow convergence 
(e.g. at very large optical depths and $\epsilon$ $\ll$ 1) lambda acceleration can shorten iteration 
times with factors of 3 to 5. For the pure scattering case ($\epsilon$=0) we however find that the lambda 
iteration can be converged within a limited number of iterations, which turn out hard to decrease further 
using the acceleration scheme. The combination of 3D source functions computed during n past iterations 
also requires to store n $\times$ $N^{3}$ additional array elements. These large memory requirements are 
a setback of the acceleration scheme when convergence is already sufficiently fast.

When 0 $<$ $\epsilon$ $\ll$ 1 however, one classically starts the source function
iterations with $\epsilon B$ \citep[see][]{Adam90} or $\sqrt\epsilon B$
\citep{Hubeny97}, but which are both useless in case $\epsilon$=0.
Towards very small values of $\epsilon$ increasingly more iteration steps $n_{\rm it}$ are
required \citep[since $n_{\rm it}$ $\sim$ 1/$\sqrt\epsilon$, e.g.][]{mihalas84},
but which do not become unacceptably large as long as the initial source function 
values are well-chosen, i.e. by starting from an escape probability approximation for 
optically thin wind conditions. For the major portion of our 3D models information 
propagates nearly directly from the stellar surface to any point in the wind, 
because there is nearly no opacity between the emission of a photon at the
stellar surface and its absorption in the wind.
The thickness of the 3D structured wind models is limited to 1 $R_{*}$ around 
the equatorial plane. The majority of the 3D wind model is therefore 
a smooth $\beta$-wind structure. Only the structured wind regions close to 
the surface (where the line optical depth increases)
slow down the convergence rate. The initial source function values in $\sim$90\% of the model
are already very close to the final values, and only $\sim$10\% of the model
(around the plane of the equator) requires further iterations of $S^{l}$ to self-consistency.

Finally, 3D lambda iteration with {\sc Wind3D} assumes symmetry about the plane of the equator of the 
central star. The code however accepts arbitrary 3D wind structures that are asymmetric about the equatorial plane.  
{\sc Wind3D} first performs the lambda iteration for the northern hemisphere, followed by the southern hemisphere.
It offers the advantage that iteration times become halved when equator-symmetric wind structures are (commonly)
adopted. Our hydrodynamic wind models are symmetric about the plane of the equator so that the lambda iteration 
can be converged for the northern hemisphere, followed with a convergence test of the source function values
in all grid-points of the southern hemisphere. 
         
\section{3D Radiative Transfer Solution}
\label{appendix 3D Radiative Transfer Solution}
An important aspect of the 3D transfer problem is that the computation of the line source function consistent
with the transfer equation through lambda iteration takes orders of magnitude more time than only solving the 
3D transfer equation (with a fully iterated source function) to compute emergent line fluxes.
This results from the time-consuming geometric and frequency integration in Eq.~(A1) required to iterate $S^{l}(\bf{p})$ 
at every grid-point, compared to the fast computation of the specific 
intensities with Eq.~(A4) when  $\chi^{l}_{ijk}$ and $S^{l}_{ijk}$ are available. The accuracy of the iterated 
line source function therefore directly determines the accuracy of line fluxes computed with the transfer equation.
This interdependence is particularly important when computing small flux variations inside wind line
profiles due to small changes of the line source function caused by local density- and velocity-structures 
in the stellar wind. In this paper we typically compute line flux variations of less 
than 100 $\rm km\,s^{-1}$ wide
in P Cygni profiles that can have total line widths of up to 5000 $\rm km\,s^{-1}$ (from the blue edge velocity 
to the red velocity end of the emission lobe). Using about 100 frequency 
points is therefore sufficient to adequately sample the line profile for solving the transfer equation. On the 
other hand, the accuracy of $\it \bar{J}_{ijk}$ in Eq.~(A5) also strongly depends on the number of frequency points
for integrating the profile function $\phi$. But to resolve small flux variations in the absorption portion of the
wind profile, the width of the line profile function has to remain sufficiently small as to not broaden them or 
render them invisible with respect to the overall flux profile of the line. When we adopt a FWHM of $\phi$ 
below 50 $\rm km\,s^{-1}$ the frequency integration in Eq.~(A5) samples the full width of line profile function 
at only three frequency points. An under-sampling of $\phi$ therefore limits the accuracy of $\bar{J}$ and 
the resulting line source function. This problem cannot be resolved by only increasing the number of frequency 
points $N_{\nu}$ for the integration of $\bar{J}$ because it linearly increases its computation time  
(per lambda iteration cycle) at the bottleneck of the 3D transfer problem. For the problem to 
remain tractable an increase of $N_{\nu}$ therefore requires to reduce the number of directions $N_{\theta}$ 
and $N_{\psi}$ to compute $\bar{J}$. An accurate integration of $\bar{J}$ over a limited number of directions
can hence only be accomplished when the number of grid-points is also limited. 

We adopt an equidistant grid with $N_{x}$=$N_{y}$=$N_{z}$=71 for the $\bar{J}$
calculation. The method we present is to linearly 3D 
interpolate the lambda-iterated line source function on a much finer grid with $701^{3}$ grid-points to solve the 
3D transfer equation. Our method of 3D interpolating $S^{l}_{ijk}$ is very efficient and proves to be sufficiently 
accurate to resolve small high-frequency flux variations in broad P Cygni line profiles. The 3D interpolation 
procedure is permitted when the length scales of the local density- and velocity-structures in the 
wind are larger than the distance between the $\bar{J}$ grid-points, or ($\delta\rho$, $\delta\bf{v}$) $>$ 
($\Delta$ x, $\Delta$ y, $\Delta$ z). This condition applies to our hydrodynamic models we compute for 
the large-scale structures in winds of massive hot stars. Note that the 3D transfer equation (A4) has 
a term  $\chi^{l}_{ijk} \, S^{l}_{ijk}$ in which $\chi^{l}_{ijk}$= $\overline{\chi^{l}_{ijk}}\,\phi$ contains 
the line profile function as well. 
Since we 3D interpolate $S^{l}_{ijk}$ on a finer grid the 
influence of the local velocity field $\bf{v}$ on the emergent specific intensity $I_{ijk}$ 
is computed also more accurately because $\phi$ is directly dependent of $\bf{v}$ (Eq.~A2) through 
the Doppler effect. 

Following Adam's method for Cartesian grids, the monochromatic flux 
$F_{\nu}$=$\int\,I_{\nu}\,\bf{n}\,{\it d}\bf{A}$ can be computed by integrating over the three 
visible surfaces $\bf{A}$ of the grid for any given direction $\bf{n}$ determined by angles 
$\theta$ and $\psi$, 

\begin{eqnarray}
F=F_{\nu}(\nu,\bf{n(\theta,\psi)})
={\it \sum_{i=1}^{N_{x}}\,\sum_{j=1}^{N_{y}}\,I_{i,j,k=1:N_{z}}(\nu,\theta,\psi)\,\Delta x\, \Delta y\, \cos(\theta)} \nonumber \\
+{\sum_{i=1}^{N_{x}}\,\sum_{k=1}^{N_{z}}\,I_{i,j=1:N_{y},k}(\nu,\theta,\psi)\,\Delta x\, \Delta z\, \sin(\theta) \, \sin(\psi) }
\nonumber \\
+{\sum_{j=1}^{N_{y}}\,\sum_{k=1}^{N_{z}}\,I_{i=1:N_{x},j,k}(\nu,\theta,\psi)\,\Delta y\, \Delta z\, \sin(\theta) \, \cos(\psi) } \,.
\end{eqnarray} 
In Eq.~(C1) $I_{i,j,k=1:N_{z}}$ are the monochromatic intensities computed at the up- and down-side surfaces ($A_{\pm z}$) of the block, 
$I_{i,j=1:N_{y},k}$ are computed at the left- and right-side surfaces ($A_{\pm y}$), and $I_{i=1:N_{x},j,k}$ at the front- 
and back-side surfaces ($A_{\pm x}$). Equation (C1) sums the intensities from the three surfaces that are visible from any direction ($\theta$,$\psi$), over projected surface elements with $\Delta\,x$=$(x_{\rm max}-x_{\rm min})$/$N_{x}$, 
$\Delta\,y$=$(y_{\rm max}-y_{\rm min})$/$N_{y}$, and $\Delta\,z$=$(z_{\rm max}-z_{\rm min})$/$N_{z}$.       
We also compute frequency integrated intensity images at the six surfaces of the grid with  
\begin{eqnarray}
I_{\pm x}=I_{i=1:N_{x},j,k}(\pm n_{x}) = \sum_{l=1}^{N_{\nu}}\,I_{i=1:N_{x},j,k}(\nu_{l},\theta=\pm \frac{\pi}{2},\psi=0)\, \Delta \nu \nonumber \\
I_{\pm y}=I_{i,j=1:N_{y},k}(\pm n_{y}) = \sum_{l=1}^{N_{\nu}}\,I_{i,j=1:N_{y},k}(\nu_{l},\theta=0:\pi,\psi=0)\, \Delta \nu  \nonumber \\
I_{\pm z}=I_{i,j,k=1:N_{z}}(\pm n_{z}) = \sum_{l=1}^{N_{\nu}}\,I_{i,j,k=1:N_{z}}(\nu_{l},\theta=0,\psi=\pm \frac{\pi}{2})\, \Delta \nu \,,
\end{eqnarray} 
for six different combinations of perpendicular angles $\theta$ and $\psi$, and where $\Delta \nu$=($\nu_{\rm max}$-$\nu_{\rm min}$)/$N_{\nu}$ is the frequency resolution of the line profile. In this paper we apply Eq.~(C1) for detailed 
3D radiative transfer modeling of spectral lines formed in the tenuous and supersonic winds of massive 
hot stars. Applications of Eq.~(C2) for 3D radiative transfer imaging will be presented elsewhere.

\clearpage
\begin{table}
\caption{Stellar and wind parameters of HD~64760 (from \citet{Kaufer+al06}).}
\label{table parameters}
\begin{tabular}{lllll}
Effective temperature       & $T_{\rm eff}$ (K)        & 24\,600 \\
Luminosity                  & $L$ (L$_{\sun}$)         & $1.55 \times 10^5$\\
Radius                      & $R$ (R$_{\sun}$)         & 22 \\
Mass                        & $M$ (M$_{\sun}$)         & 20 \\
Terminal velocity           & $v_\infty$ (km s$^{-1}$) & 1\,500 \\
Mass loss rate              & $\dot{M}$ (M$_{\sun}$ yr$^{-1}$) & $9 \times 10^{-7}$ \\
Projected rotational velocity & $v$ sin $i$ (km s$^{-1}$) & 265 \\
Equatorial angular velocity  & $\Omega$ (s$^{-1}$)       & $1.763 \times 10^{-5}$ 
(sin $i$ = 1 assumed)
\\
\end{tabular}
\end{table}

\clearpage

\clearpage 

\clearpage 

\clearpage 

\clearpage 

\clearpage 

\clearpage 

\clearpage 

\clearpage 

\clearpage 

\clearpage 

\clearpage 

\clearpage 

\clearpage

\clearpage 

\end{document}